\documentclass[fleqn,10pt]{article}
\usepackage[danish,english]{babel}
\usepackage[fleqn]{amsmath} % Til align
\usepackage{mathtools}
 \usepackage[version=3]{mhchem} % Formula subscripts using \ce{}
 \usepackage{natbib}
\usepackage{authblk}
 \usepackage{graphicx}
\usepackage{caption}
\usepackage{subcaption}
\usepackage{subfig}
\usepackage[section,above,below]{placeins}

\newcommand{\TTi}[0]{\tilde{\theta_{\rm i}}}        
\newcommand{\TTf}[0]{\tilde{\theta_{\rm f}}}   

\begin{document}
%\begin{sfrontmatter}  
\title{Simulation of waviness in neutron guides}

\author[1]{Ursula Bengaard Hansen }
\author[1]{Mads Bertelsen}
\author[2]{Erik Bergb\"ack Knudsen} 
\author[1]{Kim Lefmann}
\affil[1]{Niels Bohr Institute, University of Copenhagen, Universitetsparken 5, Denmark}
\affil[2]{Department of Physics, Technical University of Denmark, Lyngby, Denmark}
\maketitle
\begin{abstract}
As the trend of neutron guide designs points towards longer and more complex guides, imperfections such as waviness becomes increasingly important. Simulations of guide waviness has so far been limited by a lack of reasonable waviness models. We here present a stochastic description of waviness and its implementation in the McStas simulation package. The effect of this new implementation is compared to the guide simulations without waviness and the simple, yet unphysical, waviness model implemented in McStas 1.12c and 2.0.
\end{abstract}
%\end{frontmatter}

\section{Introduction}
Neutron reflecting guides are most valuable to neutron scattering science, since they transport the neutrons from the source (moderator) surroundings to a low-background region, often 20 to 100 meters away. For time-of-flight neutron instruments, also the sheer instrument length is of value, since it gives an important contribution to improving the instrument resolution.

It is common experience that the transport efficiency of neutron guides degrades with their length. This can partially be explained by multiple reflections, causing unavoidable losses even when using mirrors with almost-perfect reflectivity. However, recent designs of elliptic, parabolic, and other types of ballistic guides \cite{Boni2008, schanzer, muhlbauer_performance_2006, bertelsen_exploring_2013, cussen_multiple_2013} have been able to reduce the number of reflections dramatically. Much simulation work has been performed along these routes, and also the first physical realizations of elliptical guides have proven of great benefit \cite{ibberson_design_2009, WISH}.

Other causes of loss in neutron guide transport are imperfections of the guides, like misalignment or waviness. The effect of misalignment has to some extent
been understood for straight guides \cite{allenspach01}, and back-of-the-envelope calculations show that present-day values of waviness are unproblematic for straight guides.
However, concerns have been raised about the severity of imperfect guide conditions for complex guide shapes like the parabolic or elliptic ones. The actual relevance of this is emphasized, as these ballistic guide types are foreseen to be used for a significant part of the instruments at the European Spallation Source (ESS), with guide lengths up to 165~m \cite{ESS, kleno2012}.

%The issue of guide misalignment is presented in a recent simulation work where it is found that losses for elliptic guides with realistic values of misalignment are of the order 10\% \cite{kleno13}. 
Simulation of guide waviness has until now been hampered by the lack of a trustworthy description of waviness from individual mirrors, and simple attempts have been found to give physically invalid results \cite{kleno13} as we will discuss in the following. We here suggest an approximate model for waviness and present the implementation within the McStas package \cite{mcstas, willendrup2014mcstas}. We will present and discuss the relevant effects of waviness:
reflection angle, illumination corrections, mirror shading, and multiple reflections. Finally, we show the effect of waviness in a few realistic guide systems. Our aim with this work is to provide an effective description of the reflectivity as a function of waviness.
\FloatBarrier
\subsection{Description of reflectivity and waviness}
In most neutron ray-tracing packages, the specular reflectivity of neutron guides is modeled by a piecewise linear function \cite{mcstas, componentmanual} that depends only on the length of the neutron scattering vector, $q$:
\begin{eqnarray}
R(q) = 
\begin{cases} 
	R_0 & (q \leq Q_{\rm c}) \\ 
     	R_0 [1-\alpha (q-Q_{\rm c})] & (Q_{\rm c} < q \leq mQ_{\rm c}) ,
\end{cases}
 \label{eq:ref}    
 \end{eqnarray}
where the critical scattering vector for natural abundance Ni is $Q_{\rm c, Ni} = 0.0217$~\AA $^{-1}$. Expressions with quadratic terms in $q$ have also been used with only minor changes in performance\cite{jacobsen13,jacobsen_corrigendum_2014}.

For a perfect surface, the low-$q$ reflectivity is unity, $R_0=1$. However, roughness on length scales smaller than the neutron coherence length will reduce the reflectivity from this value, typical values lie in the range $R_0=0.990 - 0.995$.

What we here understand as waviness is a local deviation of the surface normal of
the neutron guide, on the sub-mm to cm range, larger than the neutron coherence length. At this length scale, one can assume that the neutron reflects from a single point at the surface. Typical mean waviness values (FWHM) are of the order $10^{-4}$ to $10^{-5}$~radians \cite{swissneutronics}.

In our description of waviness, we assume that the whole guide substrate and coating have the same angular deviation as the guide surface. Hence, the reflectivity function, $R(q)$, depends only on the scattering vector $q$ and is unaffected by waviness. Instead, waviness affects the angle between the neutron and the guide surface at the reflection point. Thereby it also changes the value of $q$ and the direction of the neutron after reflection - with direct consequences for the transport properties of the guide. We assume that all rays are reflected at the surface of the guide piece, effectively this means that we assume that each layer of the supermirror has the same profile as the surface.
\FloatBarrier
\section{An algorithm for waviness simulations}
As we are only interested in an average description of waviness, we require that the guide waviness can be described stochastically. In other words, we do not
need to create and store a complete description of the guide surface 
height on sub-mm scaled grid. In addition, we require the algorithm
to be scale invariant, {\em i.e.} it does not depend upon the length scale
of the waviness, only on the root-mean-square waviness value, $w$.

When discussing the waviness problem we will solely focus on the longitudinal waviness, i.e. the waviness from rotation around on the axis transverse to the neutron beam path. The transverse waviness will not affect the illumination but only contribute through a very small alteration of the beam direction, which we will ignore in this analysis. 

As an introduction, we describe the presently used, but physically wrong algorithm,
that nevertheless fulfills these requirements. We investigate by analytics and simple ray-tracing what causes the algorithm to fail, and use this as a starting point to suggest a series of improvements in order to reach an algorithm that is in correspondence with the ray-tracing results.
\FloatBarrier
\subsection{The Gaussian waviness model}
\label{sec:oldwavy}
\begin{figure}[h!]
\centering
\begin{subfigure}[b]{0.4\textwidth}
\includegraphics[width=\textwidth]{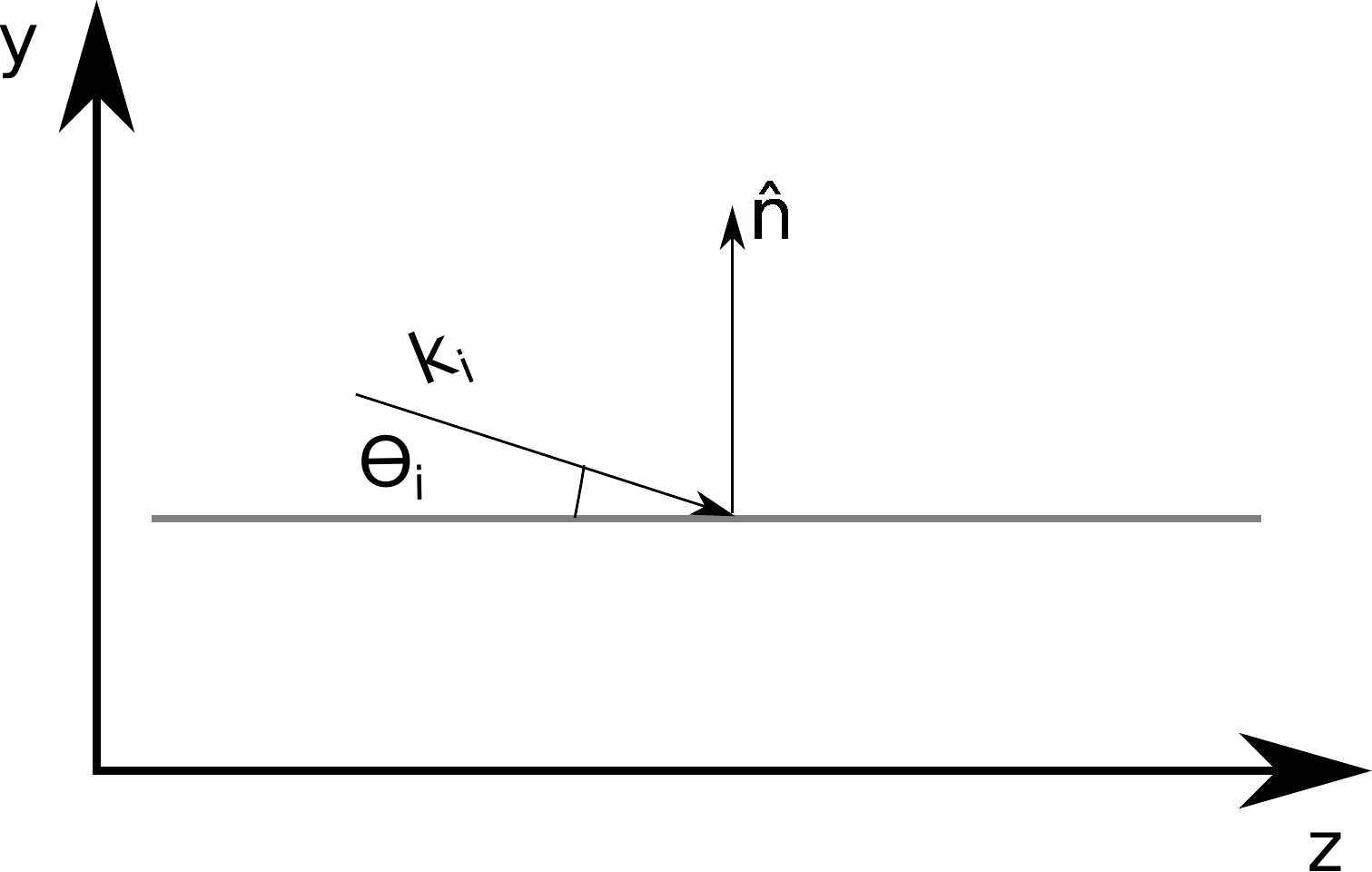}
\caption{Step 1-2.}
\end{subfigure}
\begin{subfigure}[b]{0.4\textwidth}
\includegraphics[width=\textwidth]{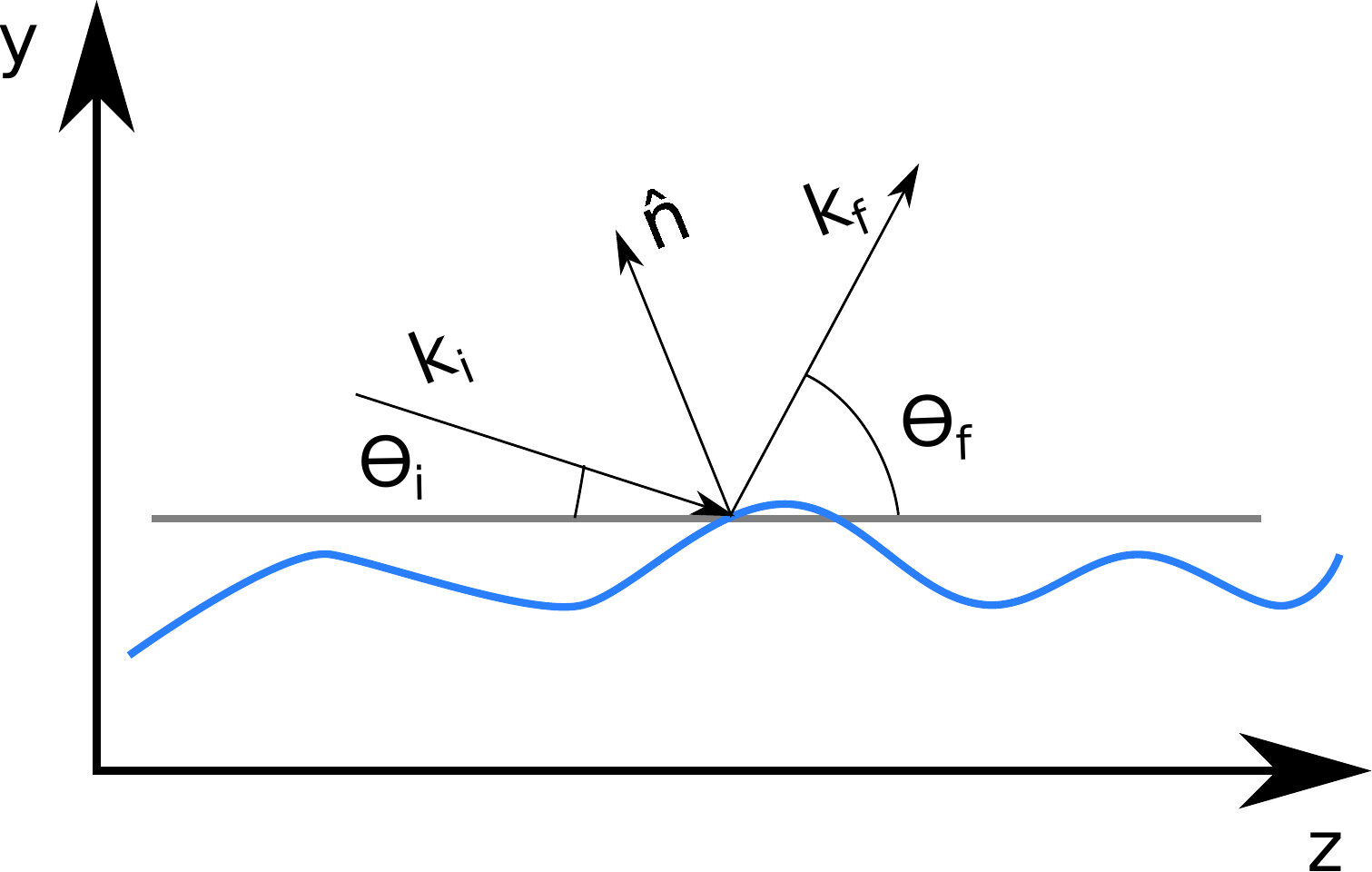}
\caption{Step 3-4.}
\end{subfigure}
\caption{Schematic drawings of the waviness algorithm implemented in \emph{Guide\_wavy}. ${\bf k_{\rm i}}$ and $\theta_{\rm i}$ (${\bf k_{\rm f}}$ and $\theta_{\rm f}$) denote the initial (final) wave vector and angle respectively. First the intersection point and the angle of incidence is found (left). Then the guide normal vector $\hat n$ is randomly modified and the exit angle $\theta_{\rm f}$ and the final wave vector is calculated. The nominal neutron direction is along $z$. The drawing is stretched along $y$ for clarity.}
\label{fig:surface_drawing}
\end{figure}

A somewhat inaccurate stochastic algorithm for the simulation of waviness 
was implemented in McStas 1.12c in the component \emph{Guide\_wavy} \cite{componentmanual}. The main steps are:
\begin{enumerate}
\item Calculate the intersection point between the neutron and the average guide surface.
\item Calculate the angle of incidence $\theta_{\rm i}$ with the average guide surface using the guide normal vector ${\bf n}$ and the direction of the incoming neutron wave vector ${\bf k}_{\rm i}$.
\item Rotate {\bf n} by a random angle, sampled from a Gaussian distribution of width $w$ to reach the local surface normal $\bf n$.
%\item Calculate the incidence angle, $\theta$, from knowledge of {\bf n} and the direction of the incoming neutron wave vector, ${\bf k}_{\rm i}$.
\item Calculate the exit angle, $\theta_{\rm f}$, and the final neutron wave vector, ${\bf k}_{\rm f}$, by requiring specular (and elastic) reflectivity of the modified surface, see Fig.~\ref{fig:surface_drawing}.
\item Calculate ${\bf q}={\bf k}_{\rm i}-{\bf k}_{\rm f}$ and $R(q)$ and use this to modify the beam intensity.
\end{enumerate}
However plausible, this algorithm yields unphysical results. The problem lies in step 3. The probability for the neutron ray to intersect the guide surface is implicitly assumed not to depend upon the local guide normal vector (from now on denoted the local waviness value). This leads to highly unphysical situations. For example, the neutron may reflect from a surface which is locally parallel to ${\bf k}_{\rm i}$. In addition it will be possible for low $\theta_{\rm i}$ values to generate negative values of $\theta_{\rm f}$.
\FloatBarrier
\subsection{Analysis of the beam illumination of a wavy surface}
\begin{figure}[h!]
\centering
 \begin{minipage}{0.49\textwidth}
\includegraphics[width=\textwidth]{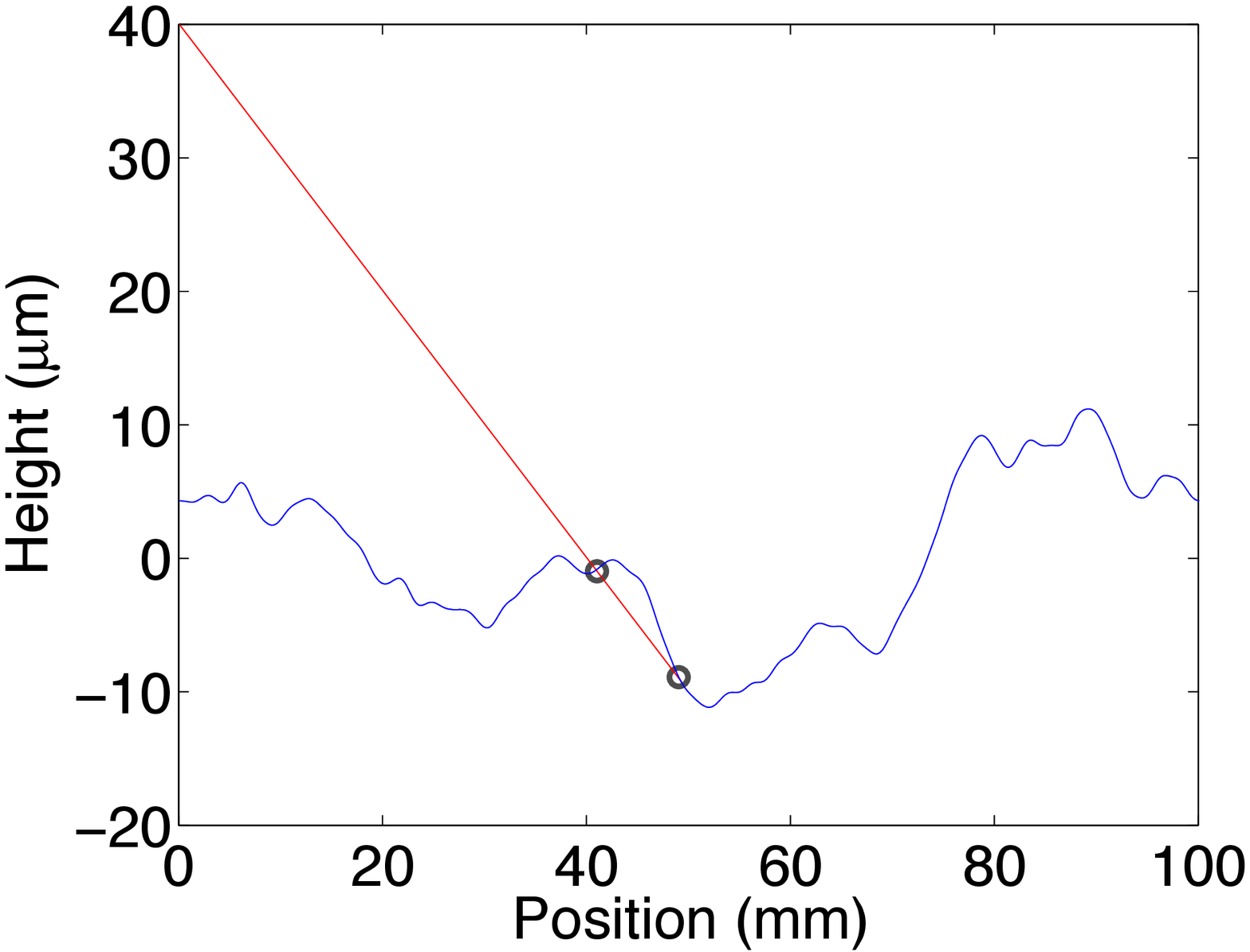}
\end{minipage}
\hfill
 \begin{minipage}{0.49\textwidth}
\includegraphics[width=\textwidth]{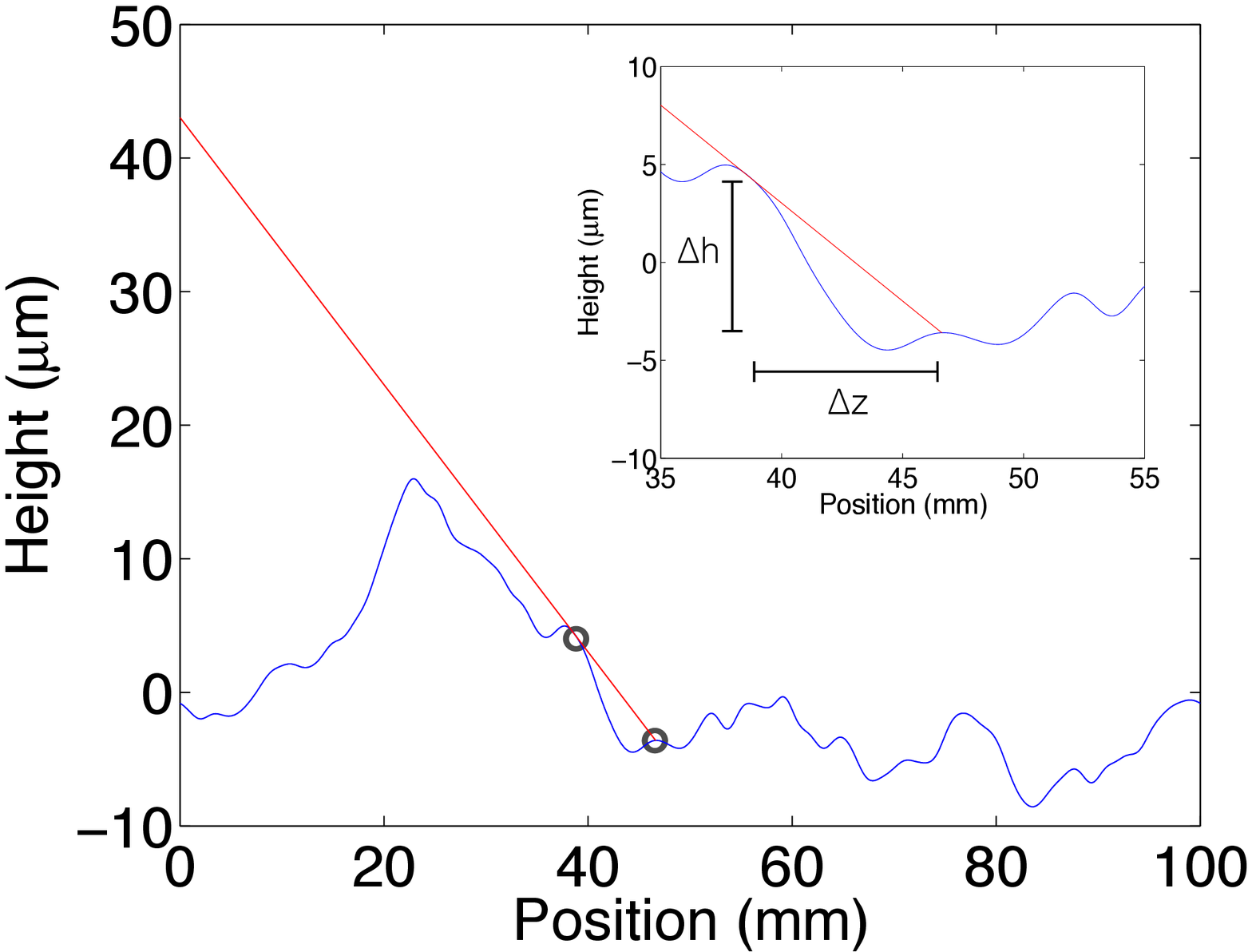}
\end{minipage}
\caption{Illustration of the problem with negative illumination and shading
in waviness simulation for a surface with $w=1$ mrad and $L=100$~mm.
We show two random, but static waviness profiles, $h(z)$ (blue), with the same low incoming neutron angle, $\theta_{\rm i} = w$ (red). Left panel shows an illegal situation where the beam at $z=41$~mm intersects the surface 'from below'. 
Right panel illustrates shading effects; the whole surface
range $z=38.6$ to 46.6~mm is out of reach for a neutron ray of the shown value of $\theta_i$.}
\label{fig:scatter_wavy}
\end{figure}
In order to arrive at a waviness algorithm based on a stochastic theorem, we start by accounting for the beam illumination. 
The total illumination of a neutron ray of nominal incidence angle $\theta_{\rm i}$ on a guide piece of length $L$ is 
\begin{equation}
\tilde f = \frac{L \sin(\theta_i)}{L}  \approx  \theta_{\rm i},
\end{equation}
as we everywhere work in the small-angle approximation. 
We now imagine that the relative height of the guide surface can be written as
$h(z)$, where we use periodic boundary conditions, $h(0)=h(L)$.
We use the McStas convention where $\hat{\bf z}$ is along the main neutron flightpath.
The local inclination is then given as
$\theta_w = \frac{dh(z)}{dz}=h'(z)$. The local illumination of this piece of guide of length $dz$
can in the small angle approximation be written as
\begin{equation} \label{eq:df}
df = dz (\theta_{\rm i} + \theta_w) .
\end{equation}
The probability $dP$ for a general neutron ray to reflect from this particular piece of guide
is its fraction of the total illumination 
\begin{equation} \label{eq:dP}
dP = \frac{df}{f} =  \frac{1 + h'(z)/\theta_{\rm i}}{L} dz.
\end{equation}
The integral of (\ref{eq:dP}) over the full length of the guide piece, $L$, gives 
\begin{equation}\label{eq:dP_int}
\int_0^L dP(z) = \int_0^L\frac{1 + h'(z)/\theta_{\rm i}}{L} dz = 1,
\end{equation}
as the integral over $h'(z)$ vanishes due to the periodic boundary conditions of $h(z)$. In addition, we
note that our model is scale invariant, as it depends only on the values
of $h'(z)$, whose magnitude is determined by the waviness $w$, and not
on the guide piece length, $L$.

\begin{figure}[h!]
\centering
 \begin{minipage}{0.49\textwidth}
\includegraphics[width=\textwidth]{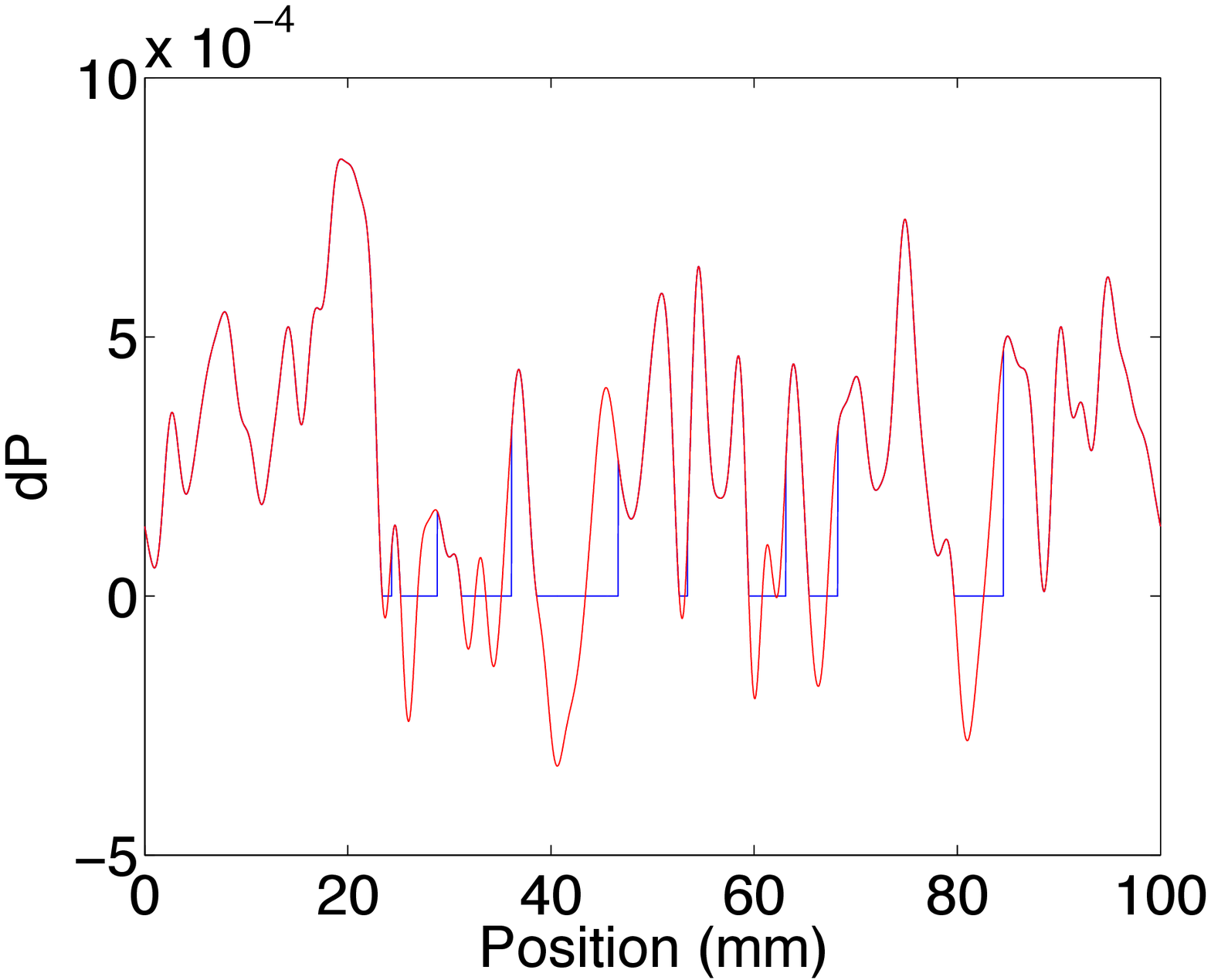}
\end{minipage}
\hfill
 \begin{minipage}{0.49\textwidth}
\includegraphics[width=\textwidth]{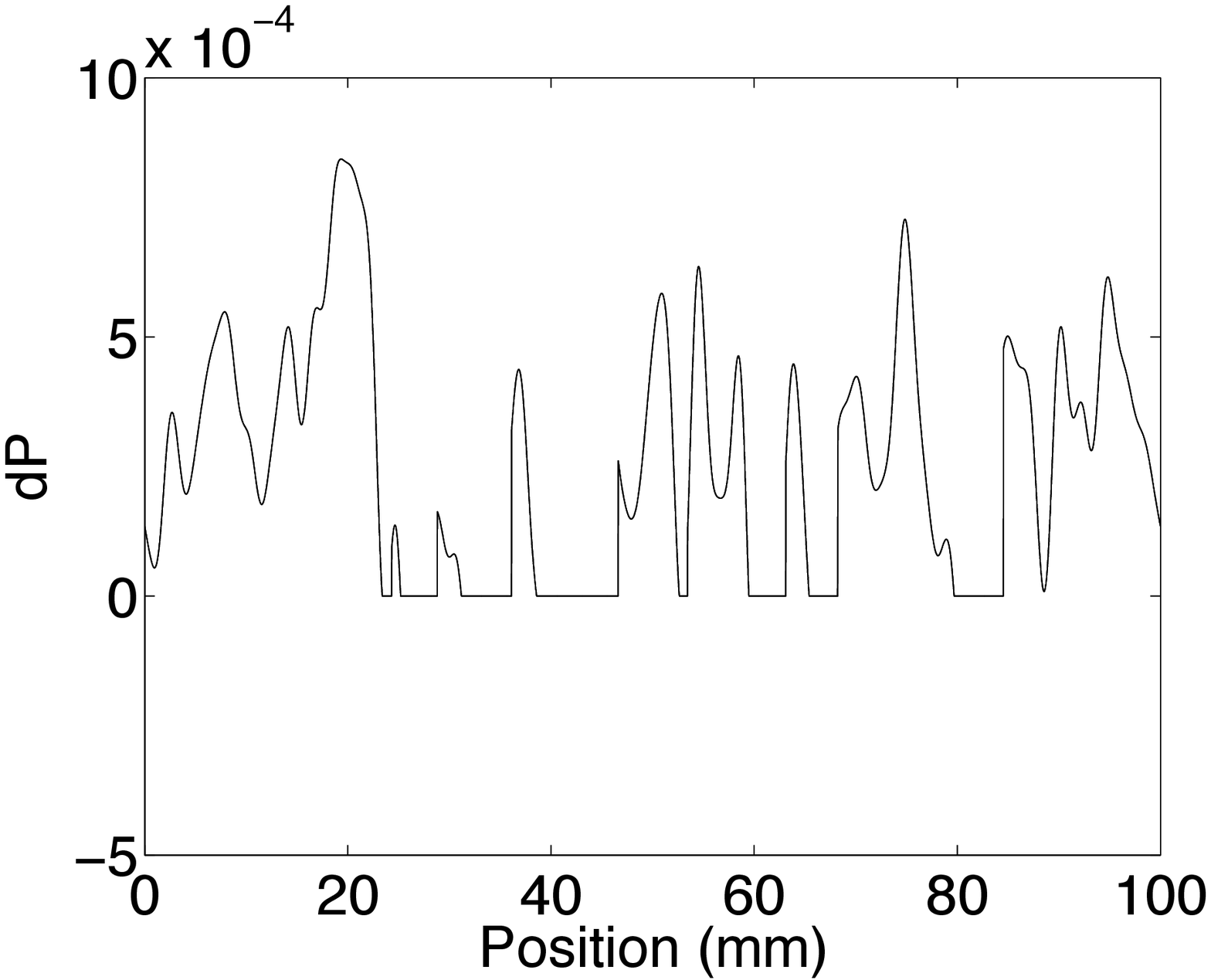}
\end{minipage}
\caption{Illustration of the influence of waviness shading on the beam illumination for a surface with $w=1$ mrad and $L=100$~mm. Left panel shows the 'bare' illumination probability (\protect\ref{eq:dP}) (red) and the shading corrections to this equation (blue) for the same surface as in Fig.~\protect\ref{fig:scatter_wavy} (right). Right panel shows the modified $dP$ value (black).
}
\label{fig:dP}
\end{figure}

A necessary requirement for describing a correct waviness reflectivity algorithm is to be able to sample the local inclination according to the probability distribution (\ref{eq:dP}).
However, for small values of $\theta_{\rm i}$, (\ref{eq:dP}) can give
the unphysical value $dP<0$. This implies that the local inclination angle of the guide
surface is higher than the incoming angle. Thus, this part of the surface cannot be reached, as illustrated in Fig.~\ref{fig:scatter_wavy}.
In fact, this implies that also other parts of the guide surface are out of reach, or ''shaded''.
Also for the shaded part of the
guide surface, the physical reflection probability must be zero, $dP = 0$. 
The figure also shows that the condition for the end of the shaded region is
$\theta_{\rm i} + \Delta h/\Delta z = 0$, corresponding to 
\begin{equation}
\int_a^b \frac{1 + h'(z)/\theta_{\rm i}}{L} dz = 0,
\end{equation}
where $a$ and $b$ are the beginning and end of the shaded region, respectively.
Hence, by replacing $dP$ by zero in the shaded region, \eqref{eq:dP_int} is still fulfilled, which was also verified numerically. 
This replacement of $dP$ is shown in Fig.~\ref{fig:dP} for the particular guide profile example shown in the right part of Fig.~\ref{fig:scatter_wavy}. Here, the shaded region is between $z=38.6$ and 46.6~mm is assigned the corrected value $dP=0$.

We choose, without loss of generality, to shift the endpoints of the periodic static surface so that the function $h(z)$ 
does not cause shading at the upper end of the interval. 
\FloatBarrier
 \subsection{A static model for random wavy surfaces}
In order to simulate the effects of shading in wavy surfaces, we must be able
to generate these surfaces. Waviness is a random
phenomenon arising when manufacturing the neutron guide surfaces. It
therefore makes sense to model waviness as a stochastic process (See for
instance~\cite{cox1965theory}) along the length of the guide surface.

Our starting assumption is that the height variations of the imperfect
surface may be written as a sum of independent stochastic processes
$h_n(z)$ with amplitudes $a_n$ and random phases $\Phi_n$:
 \begin{equation}
 \label{hz}
 h\left(z\right) = \sum_{n=1}^{n_max} h_n(z) = \sum_{n=1}^{n_{max}} a_n
\sin\left(2 \pi \frac{n}{L}z + \Phi_n\right)
 \end{equation}
which means that waviness may be expressed as:
 \begin{equation}
 \label{dhz}
 h'\left(z\right) = \sum_{n=1}^{n_{max}} 2\pi\frac{n}{L} a_n
\cos\left(2 \pi \frac{n}{L}z + \Phi_n\right)
 \end{equation}
This definition may obviously be interpreted as a Fourier series of
the surface profile.
Denoting the distribution of the phases by $f_\Phi$ the expectation, $E$, of
the part processes is:
 \begin{align}
 \label{mhn1}
 m_{h'_n}(z) = \mathrm{E}\left(h'_n(z)\right) = \int 2\pi\frac{n}{L} a_n \cos\left(2 \pi \frac{n}{L}z + \phi \right)
f_\Phi(\phi)\mathrm{d}\phi
 \end{align}
If we now assume the random phases to be uniformly distributed, the expectation of the sum
process $m_h(z)=0$.
% \begin{align}
% \label{mhn}
% m_{h'_n}(z) &=
% \int_0^{2\pi} 2\pi\frac{n}{L} a_n \cos\left(2 \pi \frac{n}{L}z +
%\phi\right) \frac{1}{2\pi}\mathrm{d}\phi  \\
%&= \frac{n}{L}a_n \int_0^{2\pi} \cos\left(2 \pi \frac{n}{L}z +
%\phi\right)\mathrm{d}\phi = 0
% \end{align}
%Obviously, as the parts are independent, 
 Similarly, the autocovariance function of the sum process \eqref{dhz} is:
 \begin{equation}
 \label{acf1}
\gamma(s,t) = \sum_{n=1}^{n_{max}} \mathrm{E} \left( h_n'(s) \right)
 \mathrm{E} \left( h_n'(t) \right) - m_{h_n'}(s) m_{h_n'}(t)
 \end{equation}
 where each of the elements in the sum is
 \begin{align}
& \gamma_n(s,t) = \int_0^{2\pi} 2\pi\frac{n}{L} a_n \cos\left(2 \pi \frac{n}{L}s + \phi\right) 2\pi\frac{n}{L} a_n \cos\left(2 \pi
\frac{n}{L}t + \phi\right) \frac{1}{2\pi}\mathrm{d}\phi \\
&=  2\pi\frac{n^2}{L^2} a_n^2 \int_0^{2\pi} \frac{1}{2} \left( \cos\left(2 \pi \frac{n}{L}(s + t) + 2\phi\right) + \cos\left(2 \pi
\frac{n}{L}(s-t)\right) \right) \mathrm{d}\phi  \\
&= 2\pi^2\frac{n^2}{L^2}a_n^2 \cos\left(2\pi\frac{n}{L}(s-t)\right)
 \end{align}
 Thus,
 \begin{equation}
 \label{acf}
 \gamma(s,t) = \frac{2\pi^2}{L^2}\sum_{n=1}^{n_{max}} n^2 a_n^2
\cos\left(2\pi\frac{n}{L}(s-t)\right) = \gamma(\tau);\ \tau=s-t
 \end{equation}
 which also proves that the sum process is stationary as the autocovariance function only depends on the distance between two samples, not on the absolute location along $z$. The stationarity property also proves our earlier statement that we may choose to shift the generated surface $h(z)$ so we avoid shading in the beginning of the considered interval.

Waviness, $w$, is generally specified as a standard deviation of the
angle variations along the length of a guide. In terms of the defined
stochastic process, this is simply $w^2=\gamma(0)$.
 What remains is to define the amplitudes (or Fourier coefficients)
of the part processes. The spectral properties hereof must depend
on the actual neutron guide manufacturing procedure. For simplicity,
here we choose
 \begin{equation}
 \label{a_n}
  a_n \propto \frac{\exp\left({-n/n_0}\right)}{n}
  \end{equation}
which indicates a fairly uniform weighting of low frequency components while suppressing high frequency variations, which are generally not significant for the purpose of describing waviness. To simulate a guide, we generate a realization of equation~\eqref{dhz} with amplitude coefficents given by~\eqref{a_n} normalized to the desired
waviness using~\eqref{acf}. It is worth to notice that this model is scale invariant and only depends on $w$, hence all profile simulations have been scaled to reasonable length scales \cite{morph}. The surface profiles in Fig. ~\ref{fig:scatter_wavy} and in the remainder of this report are generated in this way, using $n_0 = 10$ and $n_{max} = 25$. At these values the results converged. A fully stochastic description of waviness should also allow for random fluctuations in the amplitude coefficients, but as we will show in the following, the proposed model, while simple, works quite well and is a significant improvement to the schemes that have been used so far.

\FloatBarrier
\subsection{Simulation of shading effects in the static, random model}
\begin{figure}
\centering
\begin{minipage}{0.49\textwidth}
\includegraphics[width=\textwidth]{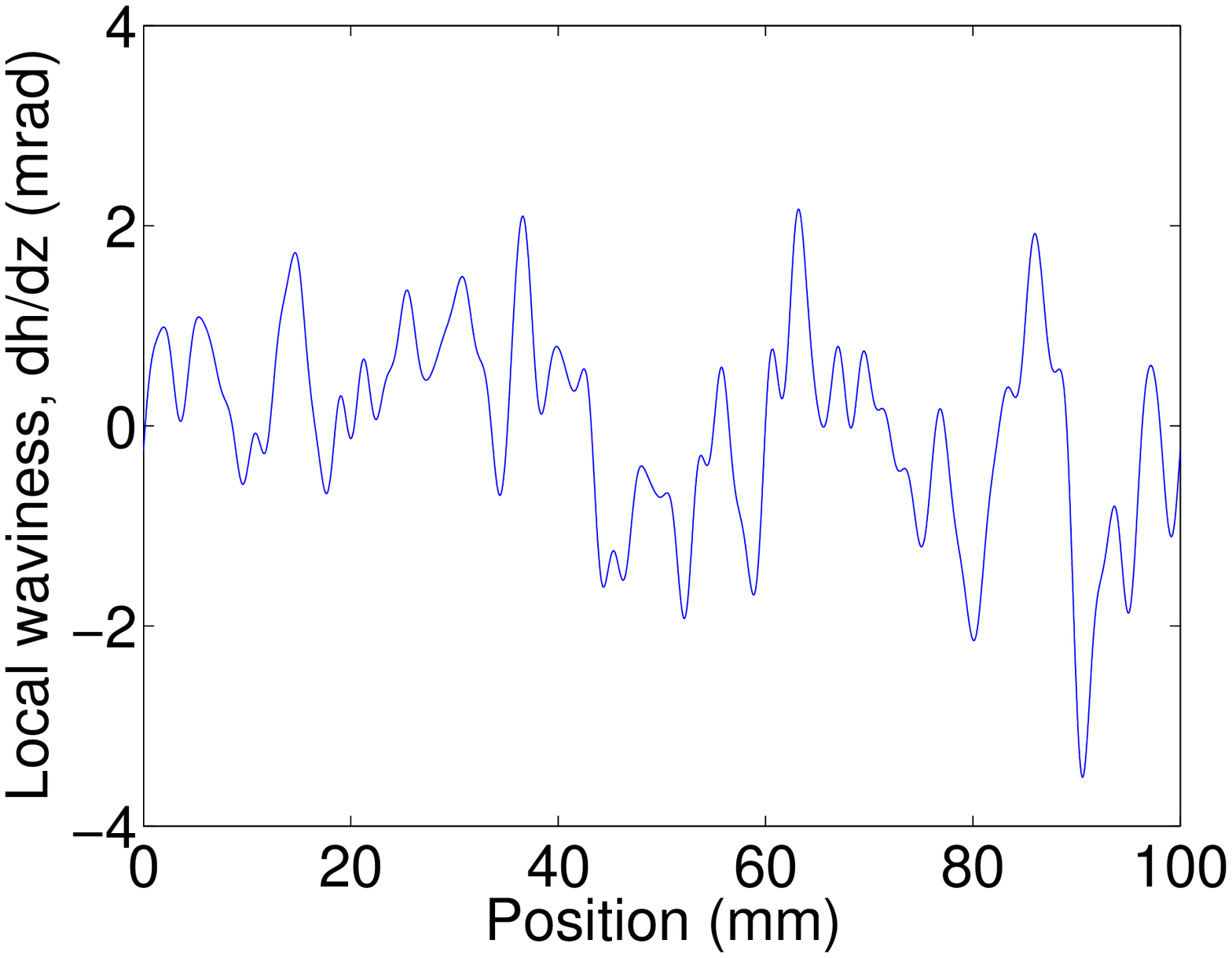}
\end{minipage}
\hfill
 \begin{minipage}{0.49\textwidth}
\includegraphics[width=\textwidth]{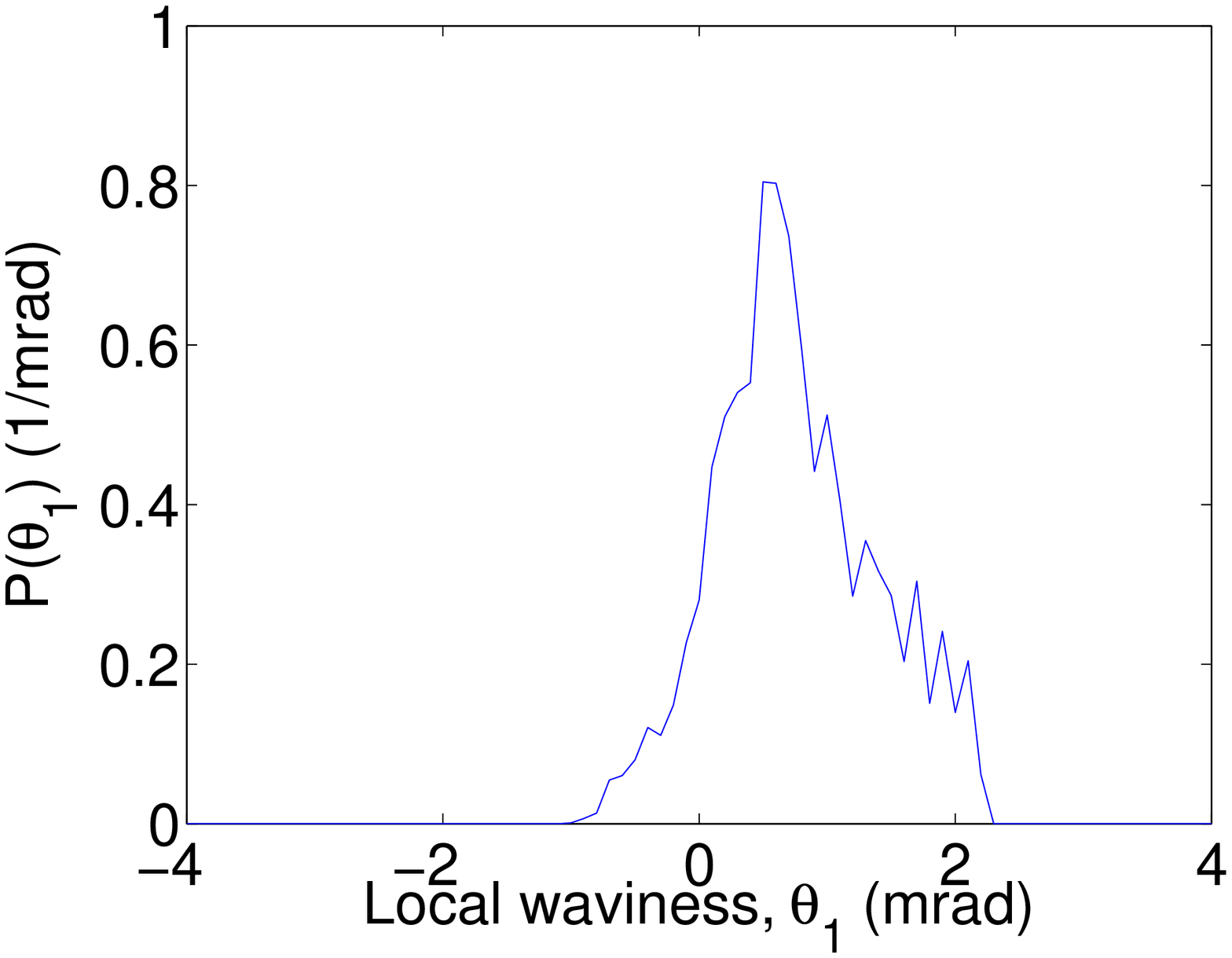}
\end{minipage}
\caption{Results from waviness shading simulation for one surface
with $w=\theta_{\rm i} = 1$ mrad.
Left panel shows the waviness profile, $h'(z)$, along the surface.
Right panel shows a weighted histogram of the local waviness value
at the point of intersection with the neutron beam.
}
\label{fig:singlesim}
\end{figure}

We have performed simulations of waviness shading on a series of surfaces. For each surface, we have calculated
$dP(z)$ and used weighted histograms to calculate the probability function $P(\theta_w)$,  describing the frequency each surface orientation is getting hit by the neutron ray. Figure~\ref{fig:singlesim} shows the result from one such simulation
with a low incidence angle, $\theta_{\rm i} = w$. The peaks in $P(\theta_w)$ corresponds to the positions where the local waviness profile is peaking. As expected from the illumination analysis, $P(\theta_w)$ vanishes
below the value $\theta_w = - \theta_{\rm i}$.

\begin{figure}[h!]
\centering
\begin{subfigure}[b]{0.49\textwidth}
\includegraphics[width=\textwidth]{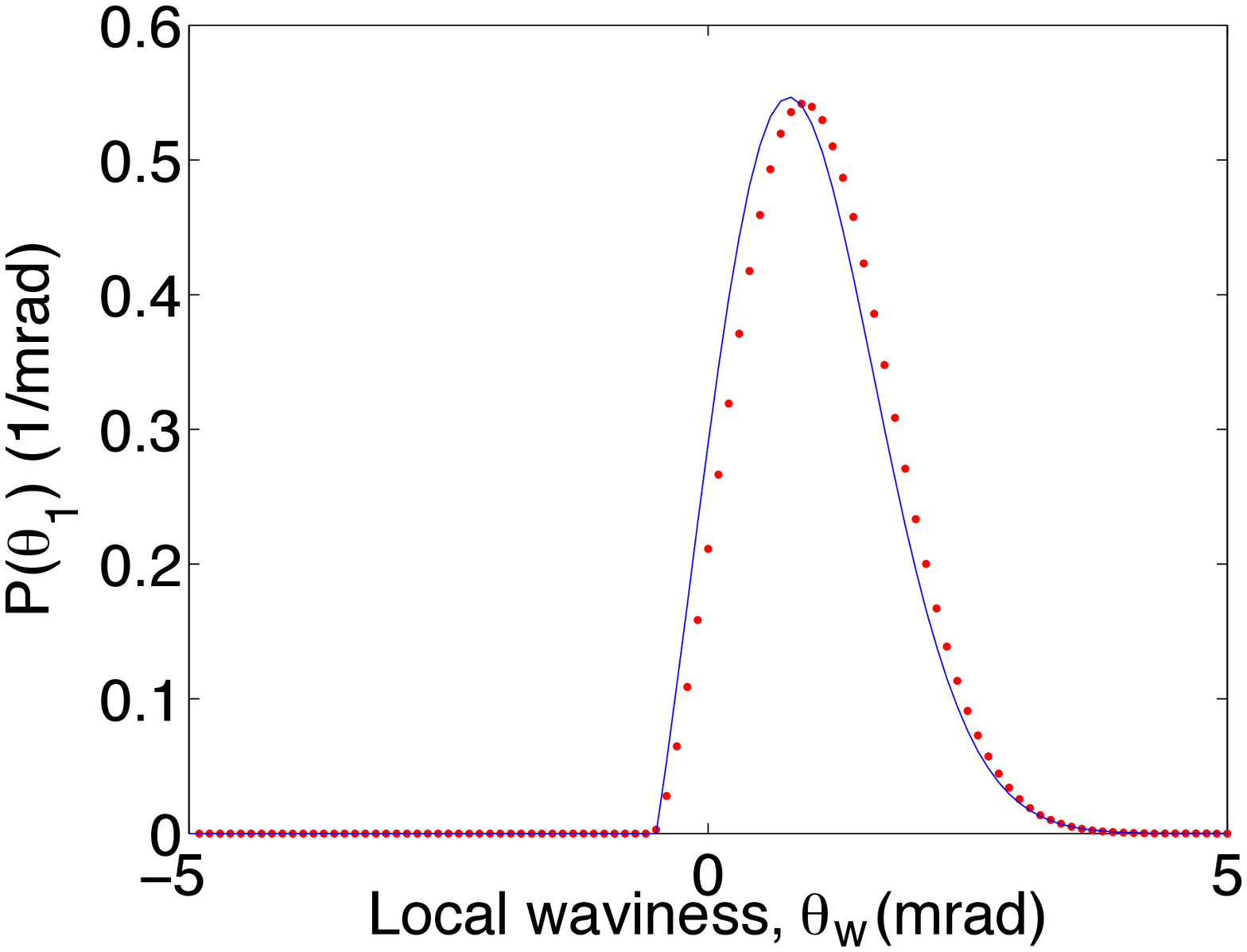}
\caption{ $\theta_{\rm i}/w = 0.5$}
\label{fig:manysim_2}
\end{subfigure}
\hfill
\begin{subfigure}[b]{0.49\textwidth}
\includegraphics[width=\textwidth]{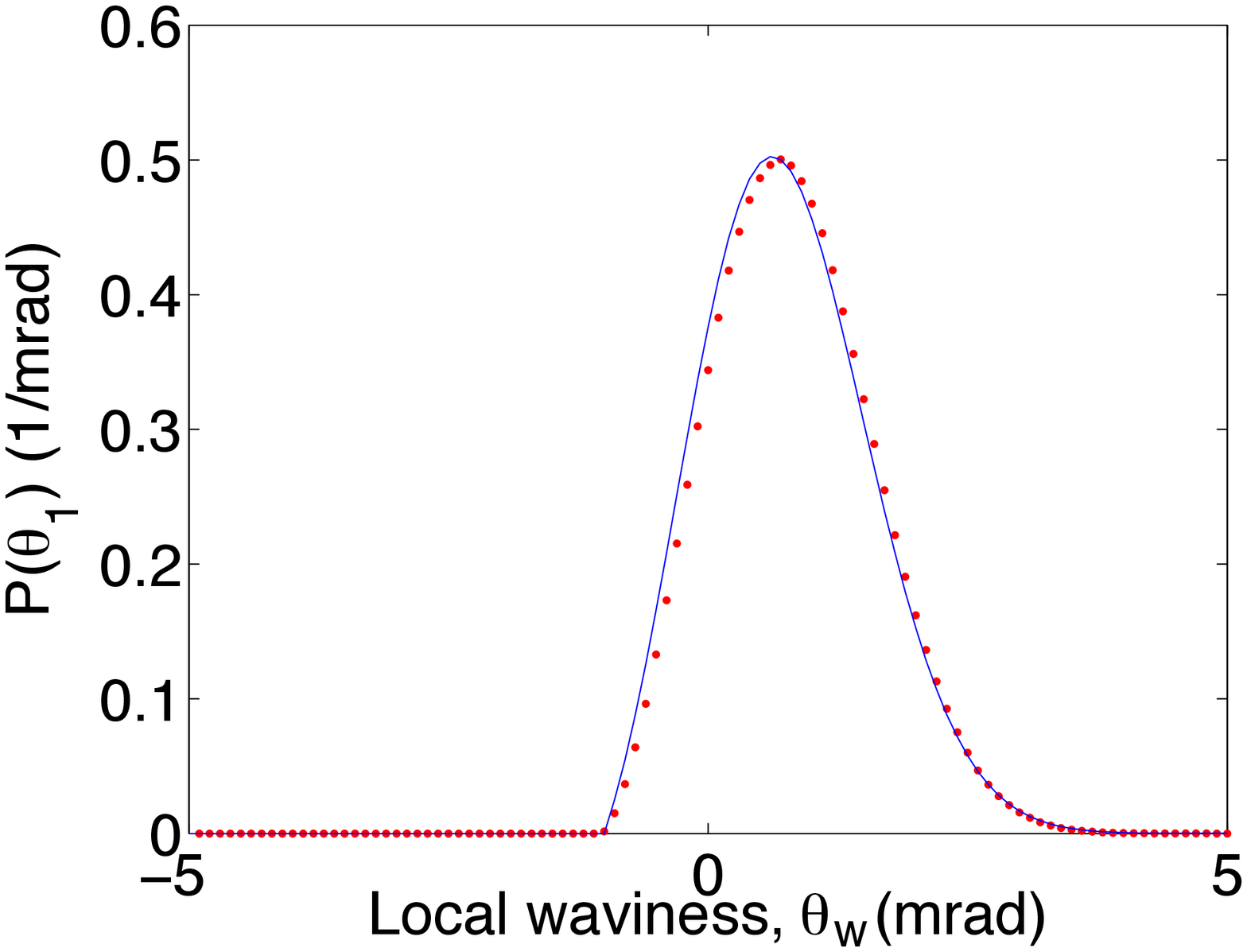}
\caption{ $\theta_{\rm i}/w = 1$}
\label{fig:manysim_1}
\end{subfigure}
\\
\begin{subfigure}[b]{0.49\textwidth}
\includegraphics[width=\textwidth]{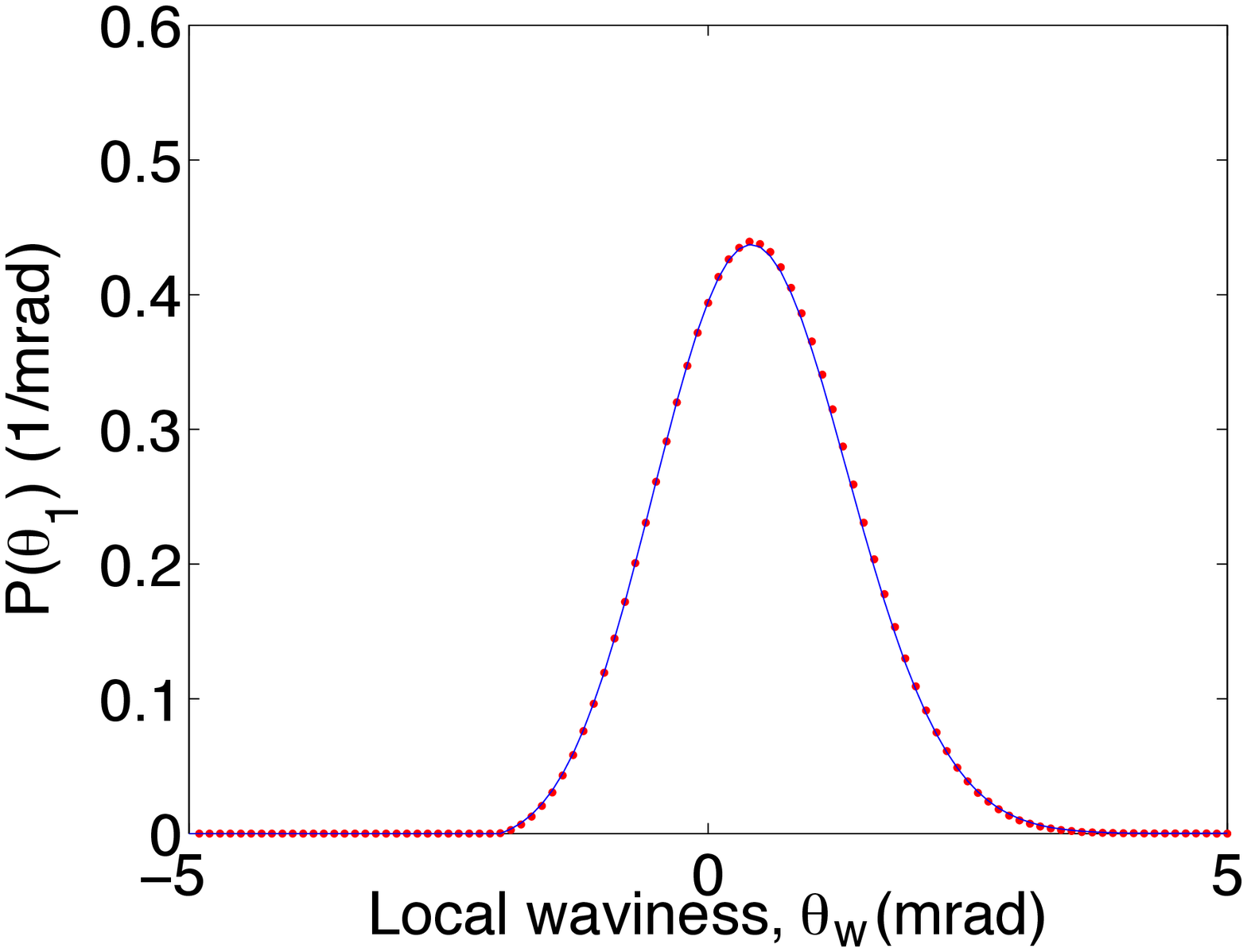}
\caption{ $\theta_{\rm i}/w = 2$}
\label{fig:manysim_3}
\end{subfigure}
\hfill
\begin{subfigure}[b]{0.49\textwidth}
\includegraphics[width=\textwidth]{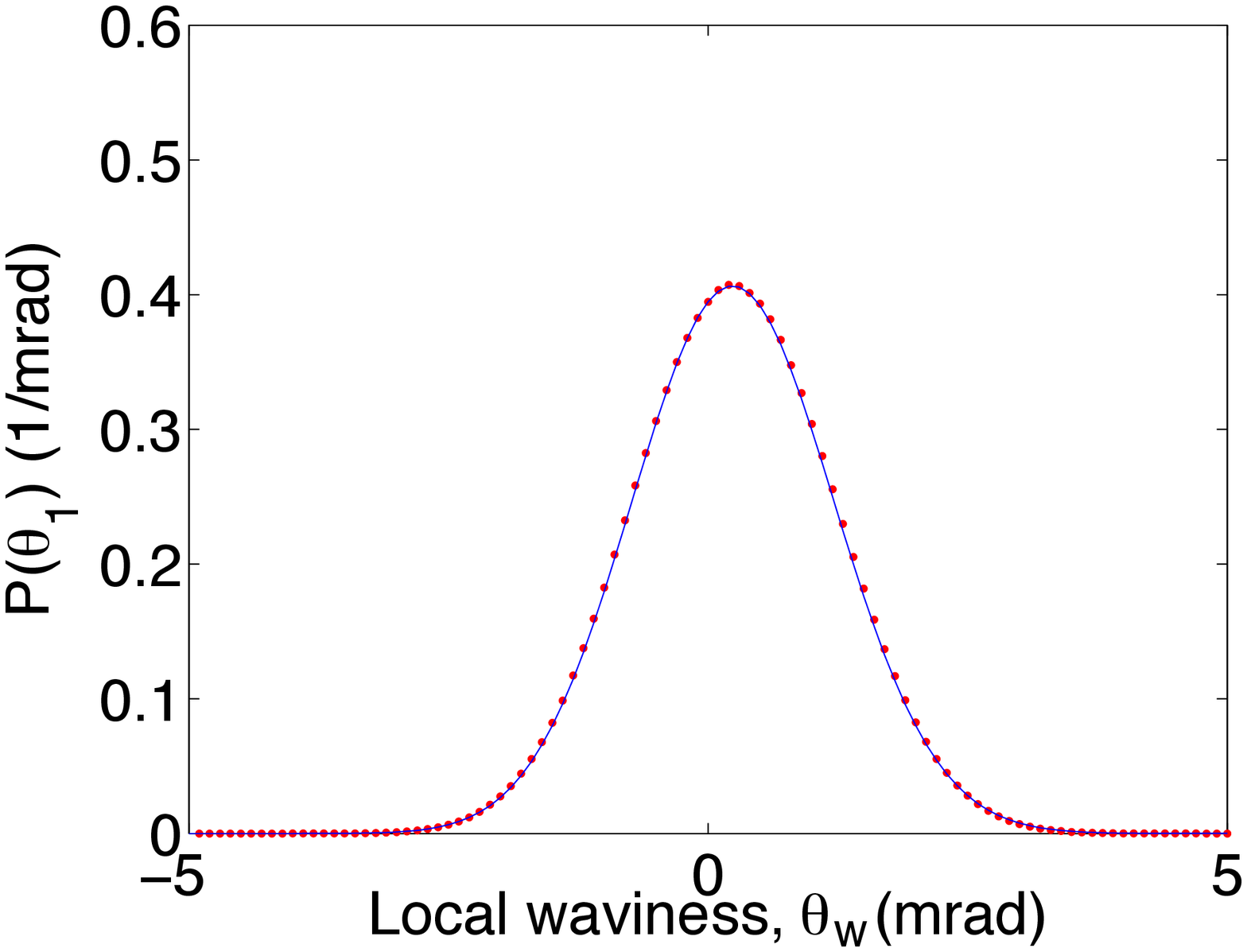}
\caption{ $\theta_{\rm i}/w = 4$}
\label{fig:manysim_4}
\end{subfigure}
\caption{Distribution of local waviness values. Dots show results from $5\cdot10^4$ waviness shading simulations for surfaces
generated with $w=1$ mrad and $\theta_{\rm i}/w = 0.5$ 1, 2 and 4. The solid line are scaled versions of eq.~(\ref{eq:ftheta}).
}
\label{fig:manysim_all}
\end{figure}

However, Fig.~\ref{fig:singlesim} shows only a single surface. For a stochastic model, all surface patterns should be represented. We simulate this by generating the average of $P(\theta_w)$ over a large number of static surfaces.
Such an average is seen in Fig.~\ref{fig:manysim_1}. We immediately
notice that the distribution vanishes at $\theta_w=-\theta_{\rm i}$
as it should, and that it otherwise seems like a skewed Gaussian with 
a center slightly larger than zero.

A simple test expression to model the simulated distribution of $\theta_w$ is found by multiplying
a normal distribution of $\theta_w(z)$ with the illumination probability
(\ref{eq:dP}):
\begin{equation} \label{eq:ftheta}
f(\theta_w) \propto dP(\theta_w) g(\theta_w,w) 
\propto  \frac{1 + h'(z)/\theta_{\rm i}}{L} \exp(-\theta_w^2 / (2 w^2)) .
\end{equation}
In figure~\ref{fig:manysim_all}, we have overlaid a scaled version of the function (\ref{eq:ftheta}) to the simulated data, and the agreement is found to be surprisingly good. For small incidence angles, $\theta_i/w=0.5$, there are small deviations at low values of $\theta_w$, where the simulated waviness is lower than the model, which in effect shifts the simulated waviness slightly to the right. This is an effect of the shading, which preferably appear at low $\theta_w$ values reducing the reflectivity. Figure~\ref{fig:manysim_all} also shows similar comparisons for higher values
of the beam incidence angle. For $\theta_{\rm i}/w=2$ or higher, the right-shift has almost vanished. This validates the picture of shading as the cause of the right-shift. We have not found a (simple) functional form of $f(\theta_w)$ that matches the simulated distribution better than eq.~\eqref{eq:ftheta}. 
\FloatBarrier
\subsection{Correction for multiple reflections}
\label{sec:sim}
Unfortunately, knowledge about the distribution of the waviness angle $\theta_w$ is insufficient to constitute a complete waviness algorithm. In addition, we must account for the neutron rays that are reflected multiple times. 
The necessity for this originates from a simple consideration: Imagine a beam
of nominal incident angle $\theta_{\rm i}$ being reflected at the local waviness angle $\theta_w$. Then, the local angle of incidence is $\theta = \theta_{\rm i} + \theta_w$, leading to an outgoing angle of
\begin{equation} \label{eq:thetaf}
\theta_{\rm f} = \theta_{\rm i} + 2 \theta_w
\end{equation}
%
%\begin{figure}[h!]
%\centering
%\includegraphics[width=0.7\textwidth]{angle_drawing.eps}
%\caption{Schematic drawing of the reflection of a neutron from a wavy surface. ${\bf k_{\rm i}}$ and $\theta_{\rm i}$ (${\bf k_{\rm f}}$ and $\theta_{\rm i}$) denotes the initial (final) wave vector and $\hat n$ the guide normal vector.}
%\label{fig:surface_drawing}
%\end{figure}
%
(See Fig. \ref{fig:surface_drawing}). For negative values of $\theta_w$ this may result in negative values of $\theta_{\rm f}$, meaning that the neutron ray continues down towards the mirror. With the exception of reflections at the very end of the mirror, there will be (at least) one other reflection before the 
neutron ray has left the mirror.

\begin{figure}[h!]
\centering
 \begin{minipage}{0.49\textwidth}
\includegraphics[width=\textwidth]{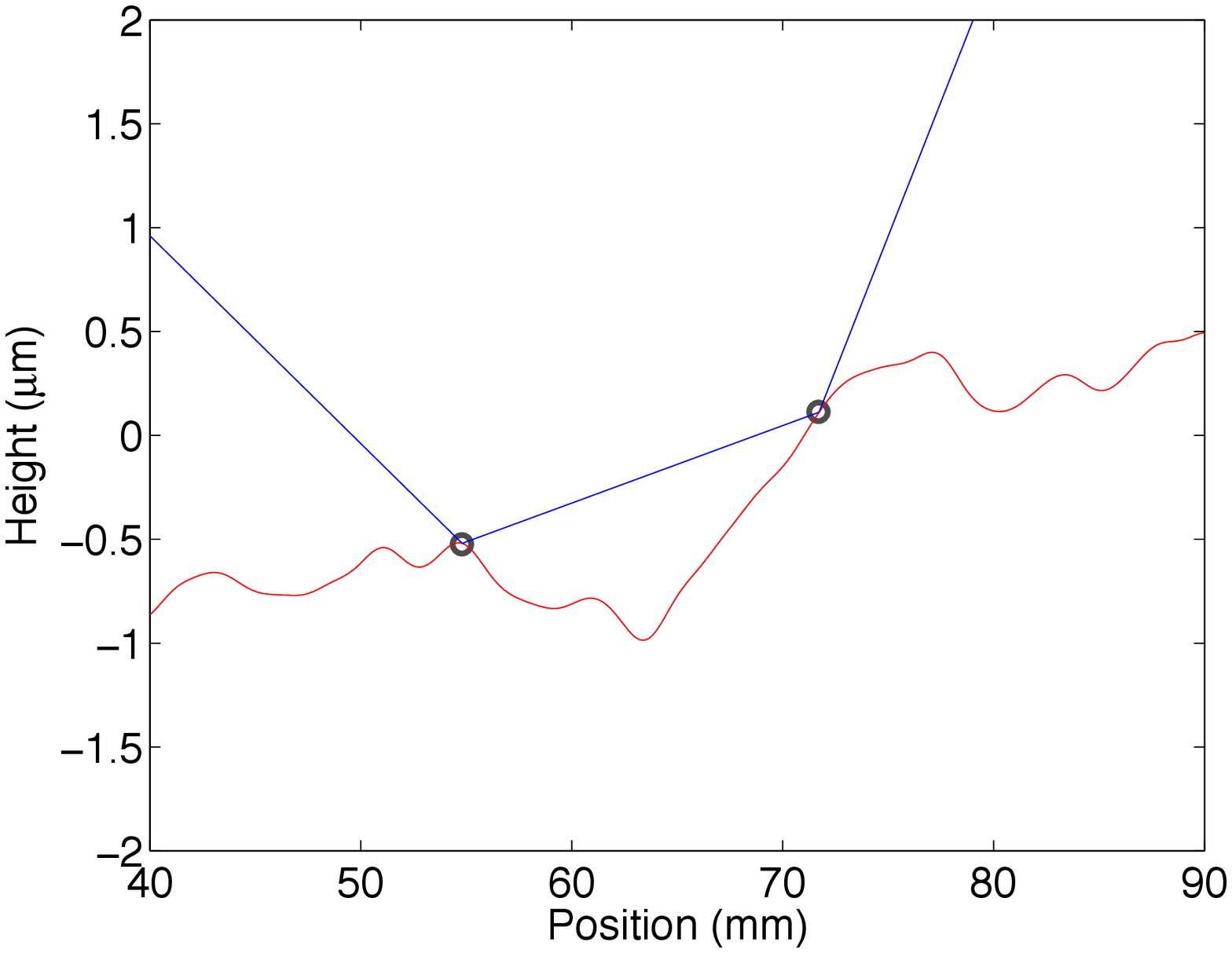}
\end{minipage}
\hfill
 \begin{minipage}{0.49\textwidth}
\includegraphics[width=\textwidth]{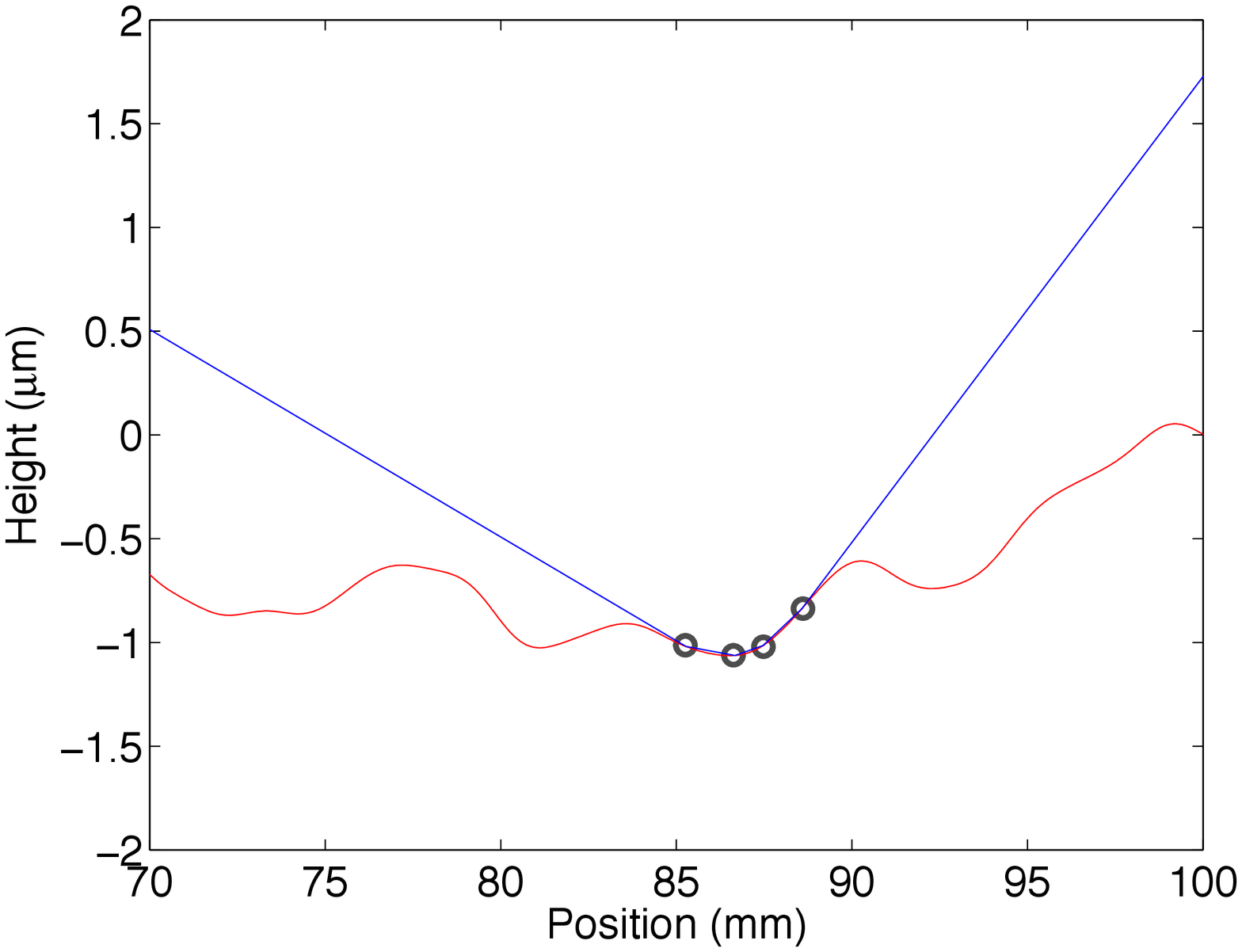}
\end{minipage}
\caption{Two examples of multiple reflections simulated with $\theta_w/w  = 1$~mrad and a repetition period $L=100$~mm. Left panel shows a typical double reflection. Right panel illustrates the rare occasion of many close multiple reflections.
}
\label{fig:multiple_ray}
\end{figure}

To quantify the effect of this, we have employed a simple ray-tracing method when simulating the wavy surfaces.
We chose the initial intersection point in accordance with the 
illumination and shading discussed above and construct the angle of the outgoing
neutron ray according to (\ref{eq:thetaf}). We then follow the outgoing ray
over an extended surface that spans over a length of $2L$.
We check across the surface height curve if the ray collides with the surface again.
If there is such a multiple collision, another outgoing angle is calculated
according to (\ref{eq:thetaf}), and the ray-tracing continues.
Two examples of these simulations are given in Fig.~\ref{fig:multiple_ray}.

\begin{figure}
[h!]
\centering
\includegraphics[width=0.7\textwidth]{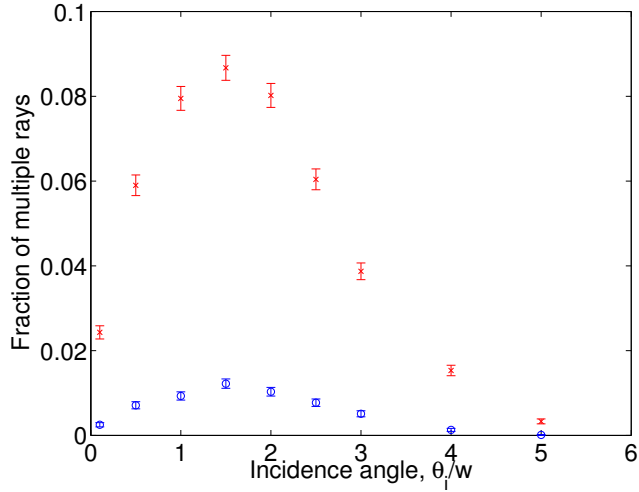}
\caption{Simulated fraction of multiple reflections as a function of incident
angle, $\theta_{\rm i}$. Red crosses represents all multiple events, while
blue circles are events with three or more reflections.
}
\label{fig:multiple_ratio}
\end{figure}

We have performed a series of simulations with varying incident angles, $\theta_i$, with $N=10^4$ simulations per angle, to obtain the fraction of multiply reflected rays. The results are shown in Fig.~\ref{fig:multiple_ratio}. We observe that the degree of multiple reflection is highest, just below 10\%, for $\theta_w/w \approx 1.5$, to decay to zero for $\theta_w/w=0$, and for 
angles higher than $\theta_w/w \approx 5$. This is in agreement with expectations, since rays with very low angles will reflect only from the tops of the height curve, $h(z)$, on the side with $\theta_w>0$ due to shading, while high incident angles will reflect with high outgoing angles without chance for having a second reflection, $\theta_i>2|\theta_w|$, leaving eq. \eqref{eq:thetaf} always positive. Fig.~\ref{fig:multiple_ratio} also shows that the ratio of rays with more than two reflections
in general behaves as the ratio of all multiple reflections. The maximum is of the order 1\%, occurring also around $\theta_w/w \approx 1.5$.

Our most important result, however, is the distribution of final reflections angles, $P(\theta_{\rm f})$.  A naive prediction of the shape of this simulation would come from taking only the illumination argument into account, ignoring shading and
multiple scattering. Here we combine
(\ref{eq:ftheta}) and (\ref{eq:thetaf}) to reach
\begin{equation} \label{eq:fthetaf}
f(\theta_{\rm f}) \propto \exp(-(\theta_{\rm f}-\theta_{\rm i})^2/(8w^2)) \left(1+\frac{(\theta_{\rm f}-\theta_{\rm i})}{2 \theta_{\rm i}}\right) .
\end{equation}

\begin{figure}[h!]
\centering
\begin{subfigure}[b]{0.49\textwidth}
\includegraphics[width=\textwidth]{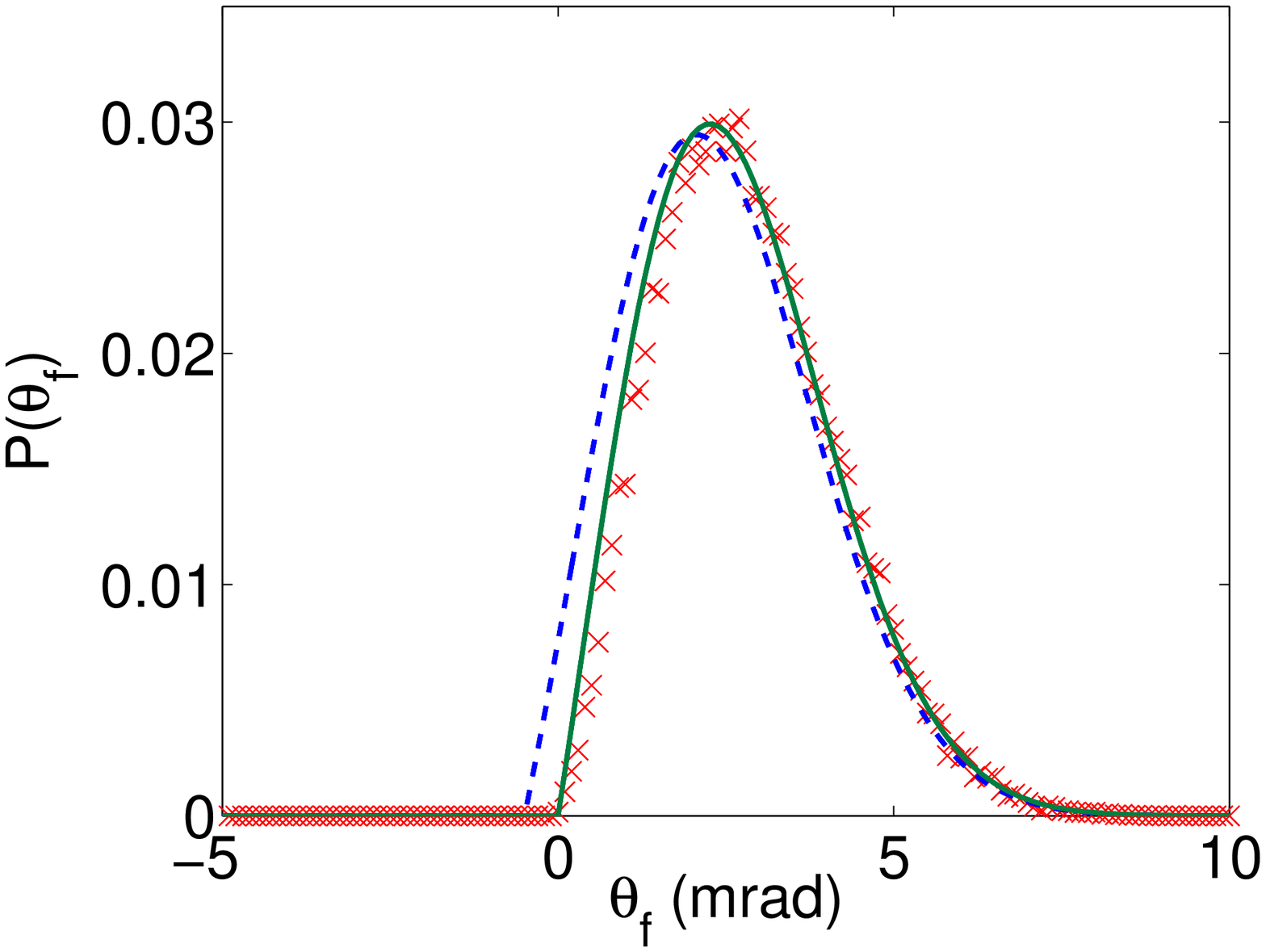}
\caption{ $\theta_{\rm i}/w = 0.5$}
\label{fig:thetaf_2}
\end{subfigure}
\hfill
\begin{subfigure}[b]{0.49\textwidth}
\includegraphics[width=\textwidth]{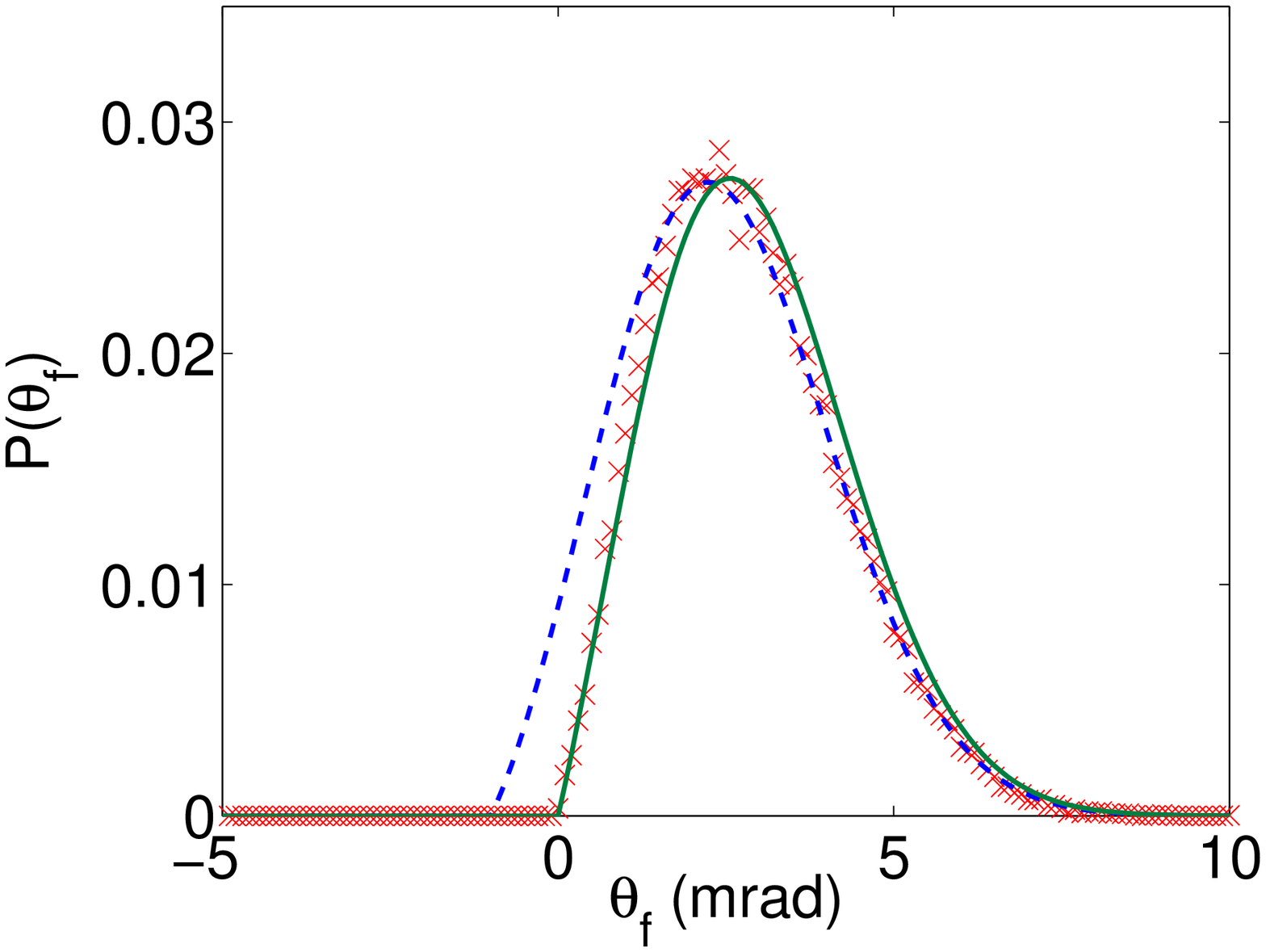}
\caption{ $\theta_{\rm i}/w = 1$}
\label{fig:thetaf_1}
\end{subfigure}
\\
\begin{subfigure}[b]{0.49\textwidth}
\includegraphics[width=\textwidth]{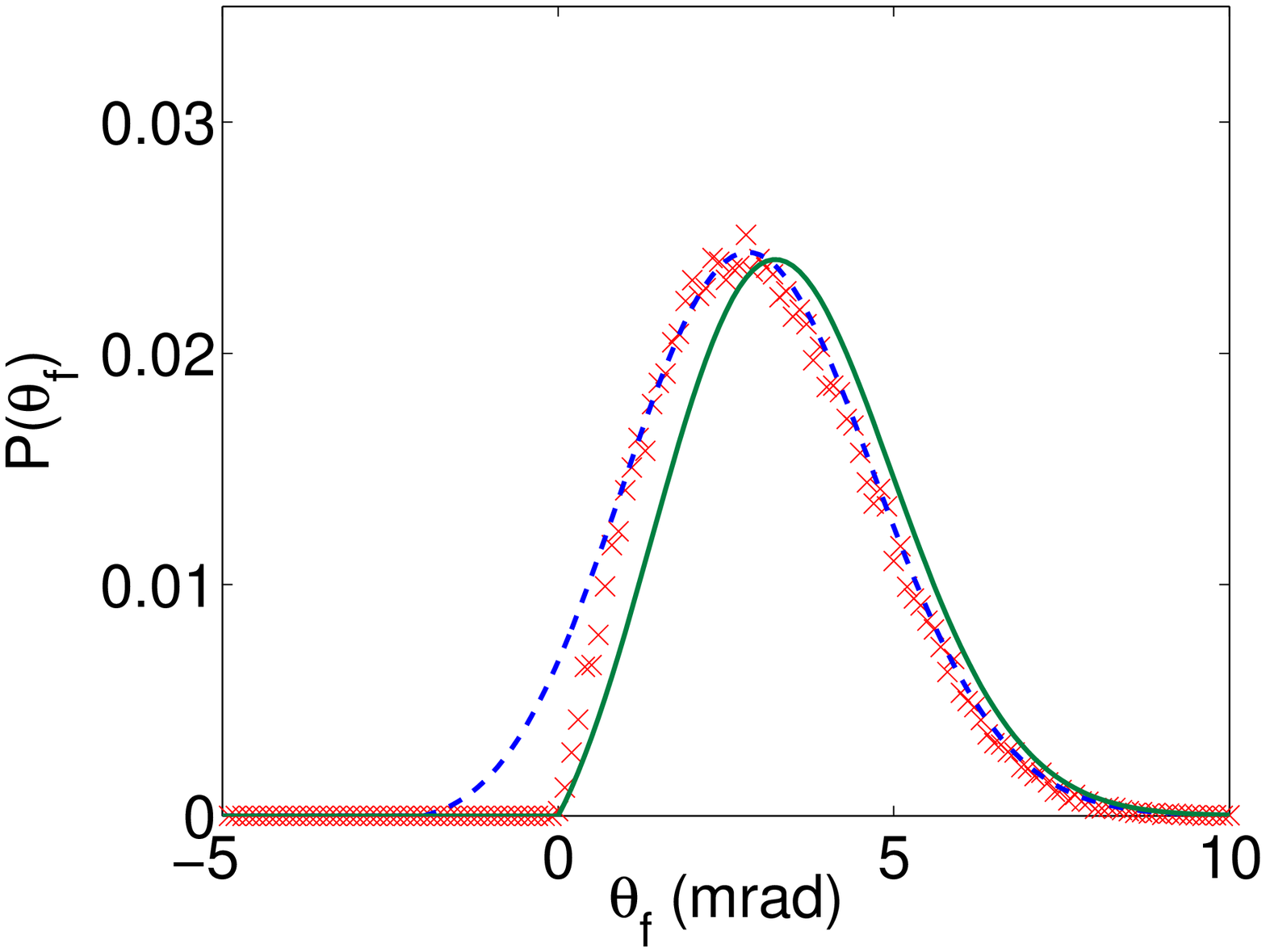}
\caption{ $\theta_{\rm i}/w = 2$}
\label{fig:thetaf_3}
\end{subfigure}
\hfill
\begin{subfigure}[b]{0.49\textwidth}
\includegraphics[width=\textwidth]{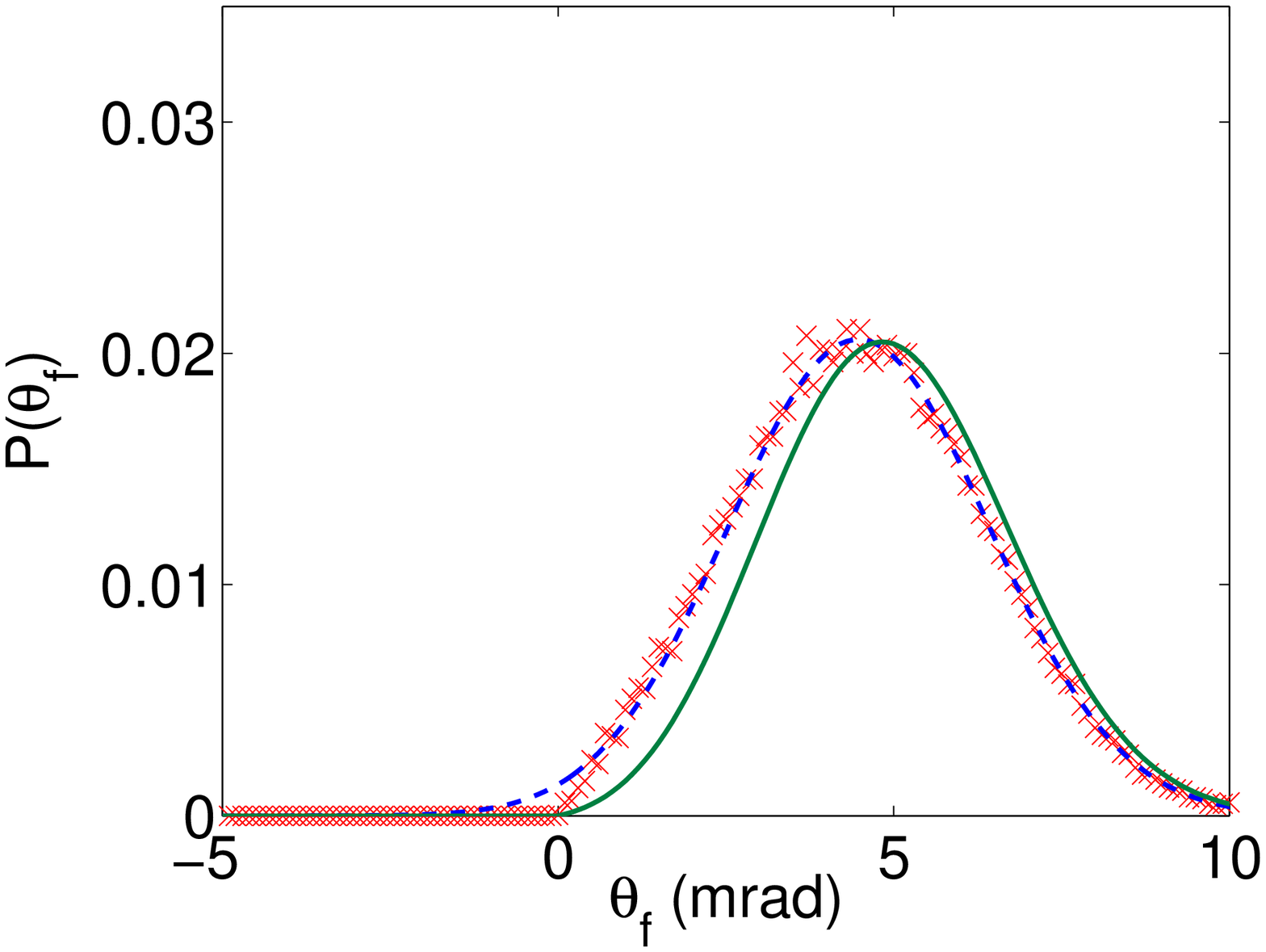}
\caption{ $\theta_{\rm i}/w = 4$}
\label{fig:thetaf_4}
\end{subfigure}
\caption{Distribution of outgoing angles of neutron rays, $\theta_{\rm f}$,  reflected from a wavy surface with $w=1$~mrad, $\theta_{\rm i}/w = 0.5, 1, 2$, and 4.  Red crosses represents results from simulations with $N=5\cdot10^4$ rays, 
the blue dashed line is the prediction (\ref{eq:fthetaf}) taking only the illumination into account, while the green line is our conjecture, (\protect\ref{eq:conjecture}).
}
\label{fig:thetaf_all}
\end{figure}

Fig.~\ref{fig:thetaf_1} shows the simulated distribution overlaid with the eq. (\ref{eq:fthetaf}). We observe that the simulations show a vanishing probability for a reflection at
$\theta_{\rm f}=0$, as one must require, while the naive prediction (\ref{eq:fthetaf})
has a finite probability at negative $\theta_{\rm f}$ values, as it does not include multiple reflections.

The full analytical description of the illumination, shading, and multiples into one equation is a complex task that we have no intention of performing.
However, we have performed a minimal change of  (\ref{eq:fthetaf})
to make it vanish at $\theta_{\rm f} = 0$:
\begin{equation} \label{eq:conjecture}
f(\theta_{\rm f}) \propto \exp(-(\theta_{\rm f}-\theta_{\rm i})^2/(8w^2)) \left(\frac{\theta_{\rm f}}{2 \theta_{\rm i}}\right) .
\end{equation}
Fig.~\ref{fig:thetaf_all} show that this equation in fact describes the simulated
distribution of the outgoing angle $\theta_{\rm f}$ quite well for $\theta_{\rm i}/w=1$. 
Hence, we use (\ref{eq:conjecture}) as a first order conjecture for the true outcome of the reflection from wavy surfaces.

We have as well performed simulations of other values of 
$\theta_{\rm i}/w$: 0.5, 2, and 4. All these results are shown along with
the predictions (\ref{eq:fthetaf}) and (\ref{eq:conjecture}) in Fig.~\ref{fig:thetaf_all}. We see that for the high angles of incidence, $\theta_{\rm i}$, the naive prediction (\ref{eq:fthetaf}) in general works better than the conjecture (\ref{eq:conjecture}).
\FloatBarrier
\section{A new approximate algorithm for waviness simulation}
\label{sec:newwavy}
Our objective is now to find a functional form that can effectively describe $P(\theta_{\rm f})$  as a function of $\theta_{\rm i}$ and $w$. From our simulations we see that $P(\theta_{\rm f})$ approaches zero for $\theta_{\rm f}\rightarrow0$. For small $\theta_{\rm i}$ it increases linearly with $\theta_f$ to a maximum slightly above $\theta_{\rm i}$ and then decreases in a Gaussian manner. In the other end of the spectrum, when $\theta_{\rm i}$ is large, $P(\theta_{\rm f})$ can be described by a Gaussian with a width $w$ centered at $\theta_{\rm i}$. Both observations are in accordance with the simple considerations discussed earlier. We have made no attempt to understand the details the shape of $P(\theta_{\rm f})$ in the intermediate region. 

As a starting point we have modified the expression \eqref{eq:fthetaf} in order to reproduce the observed behavior. We have produced three alternatives: 
\begin{align}
f_1(\TTf) &= \alpha_1 \exp(-(\TTf - \kappa_{1}\TTi)^2/8) \left(1+\frac{(\TTf - \TTi)}{2 \TTi}\right) \label{eq:f1}\\
f_2(\TTf) &= \alpha_2 \exp(-(\TTf - \kappa_{2}\TTi)^2/8) \left(\frac{\TTf}{2 \TTi}\right) \label{eq:f2}\\
f_3(\TTf) &= \alpha_3 \exp(-(\TTf - \kappa_{3}\TTi)^2/8)  \tanh(\beta_3\TTf/\TTi) \label{eq:f3}
\end{align}
where $\TTi$ and $\TTf$ are the dimensionless variables $\TTi=\theta_i/w$ and $\TTf=\theta_f/w$ respectively.  

\begin{figure}[h!]
\centering
\includegraphics[width=0.7\textwidth]{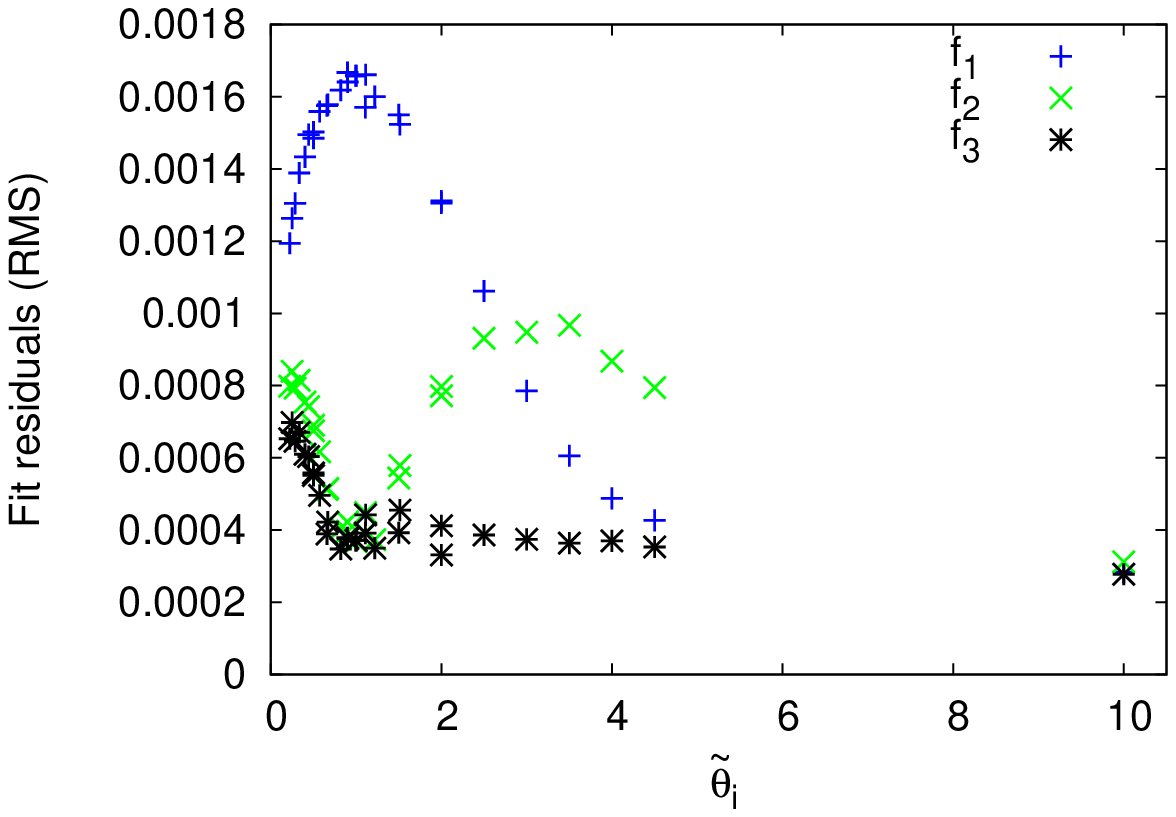}
\caption{The root mean square of the fits to $f_{1}$ (blue $+$), $f_{2}$ (green $\times$) and $f_{3}$ (black $\mathrlap{+}\times$) as a function of incoming angle $\TTi$.}
\label{fig:RMS}
\end{figure}

\begin{figure}[h!]
\centering
\begin{subfigure}[b]{0.49\textwidth}
\includegraphics[width=\textwidth]{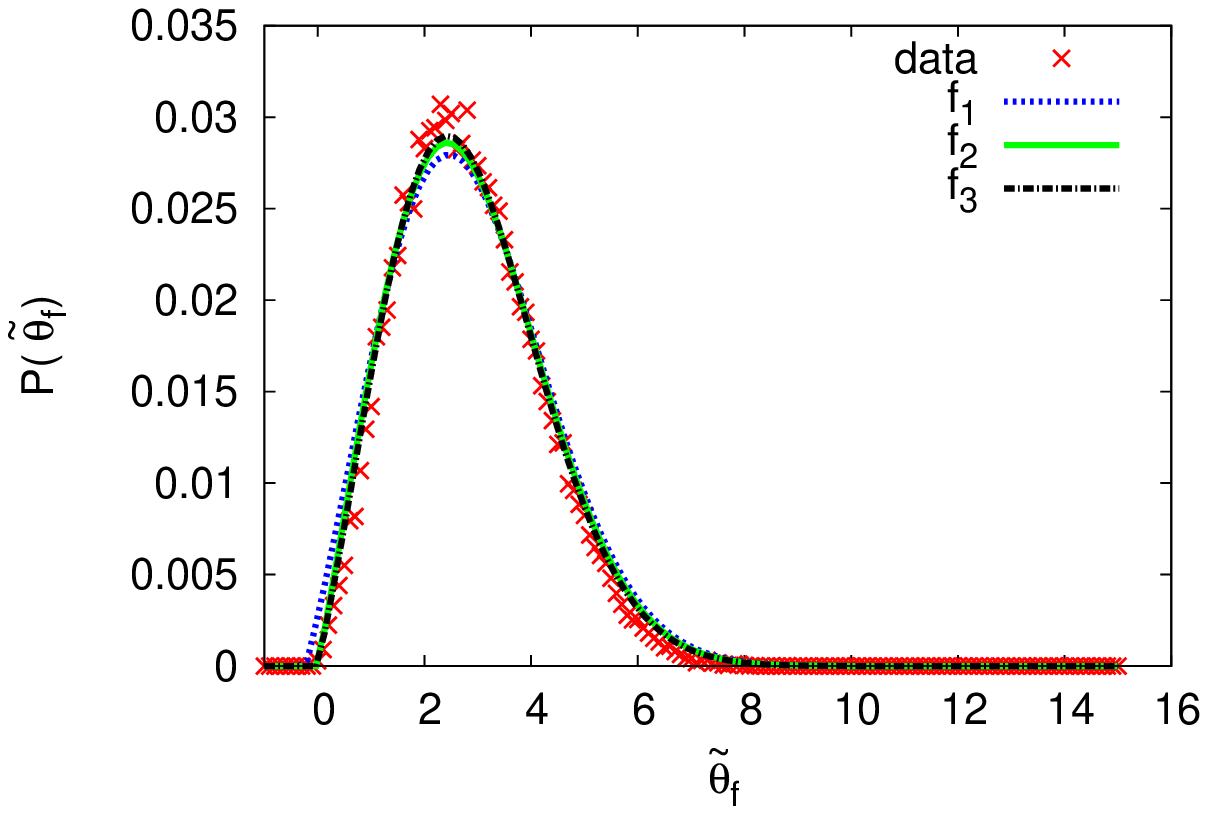}
\end{subfigure}
\hfill
\begin{subfigure}[b]{0.49\textwidth}
\includegraphics[width=\textwidth]{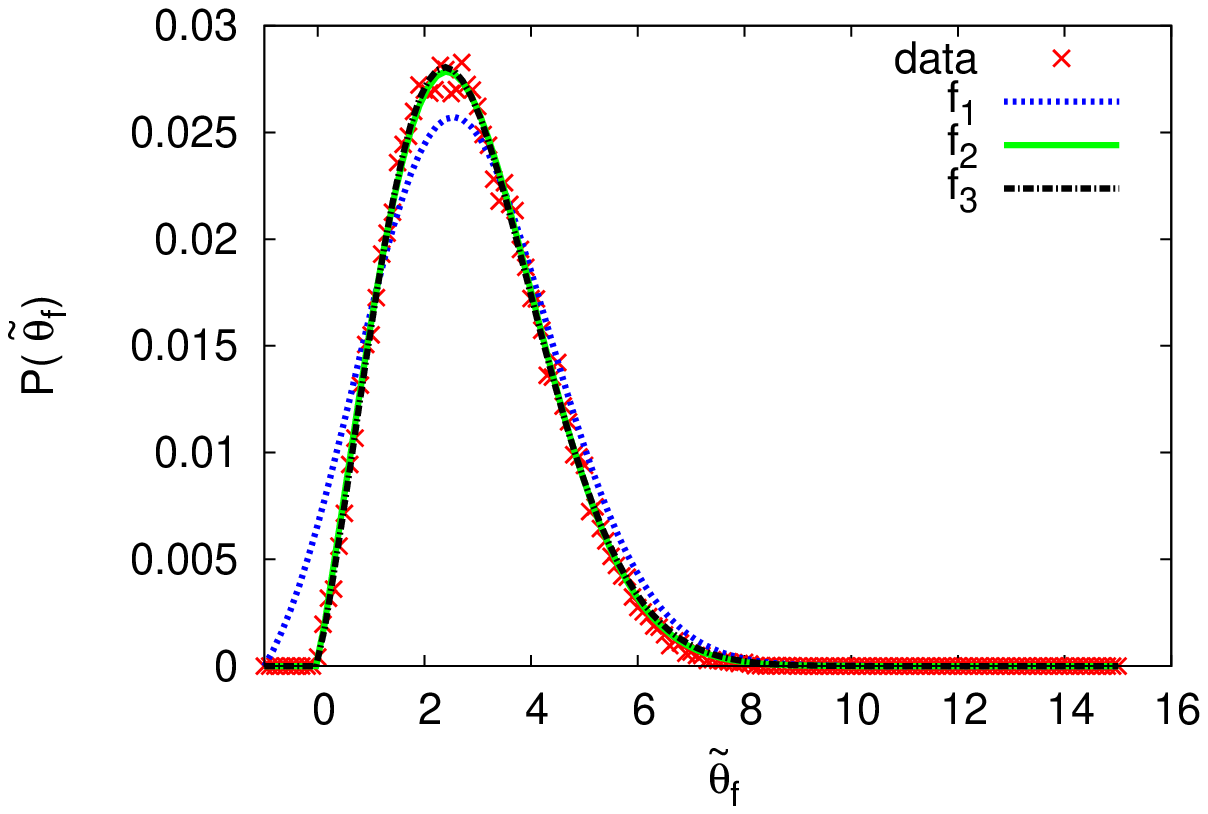}
\end{subfigure}
\\
\begin{subfigure}[b]{0.49\textwidth}
\includegraphics[width=\textwidth]{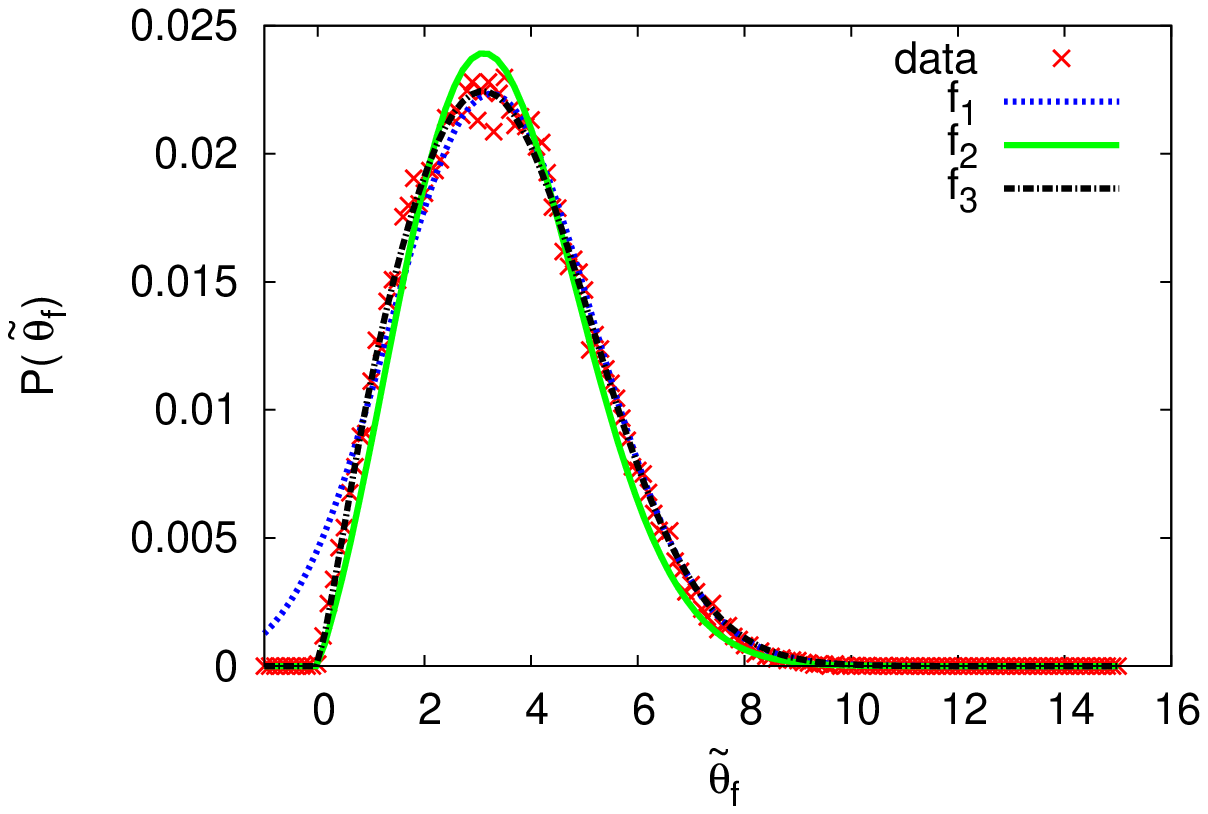}
\end{subfigure}
\hfill
\begin{subfigure}[b]{0.49\textwidth}
\includegraphics[width=\textwidth]{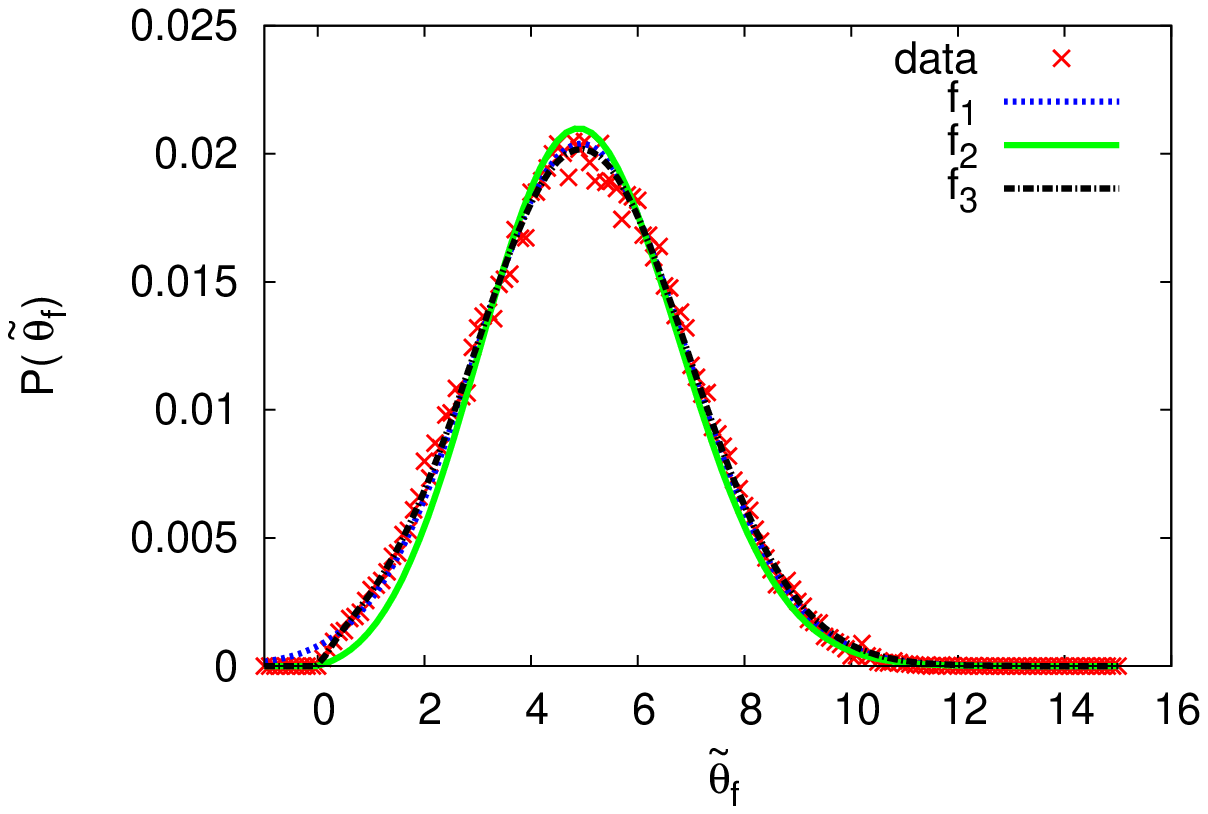}
\end{subfigure}
\caption{Simulated data for $\TTi =0.22$, 1, 2.5 and 4.5 with $N=5\cdot10^4$ rays along with fits to $f_{1}$ (blue dashed line), $f_{2}$ (green solid line) and $f_{3}$ (black dot dashed line).}
\label{fig:fit_ex1}
\end{figure}

A series of simulations with $\TTi$ going from 0.22 to 10 was made and each simulation was fitted to the expressions \eqref{eq:f1}, \eqref{eq:f2} and \eqref{eq:f3}. In all cases the $\alpha_{1,2,3}$, $\beta_{1,2,3}$ and $\kappa_{1,2,3}$ are fitting parameters that are free to vary. Fig. \ref{fig:RMS} shows the root mean square of the residuals for each fit. It is clear that the expression $f_{3}$ has the best performance over the whole range.  This can also be seen in the individual simulations, in Fig. \ref{fig:fit_ex1} examples where $\TTi =0.22$, 1, 2.5 and 4.5 are shown.  We therefore choose $f_3(\TTf)$ as our model for describing $P(\theta_f)$ when simulating waviness.
\FloatBarrier

\begin{figure}[h!]
\centering
 \begin{minipage}{0.49\textwidth}
\includegraphics[width=\textwidth]{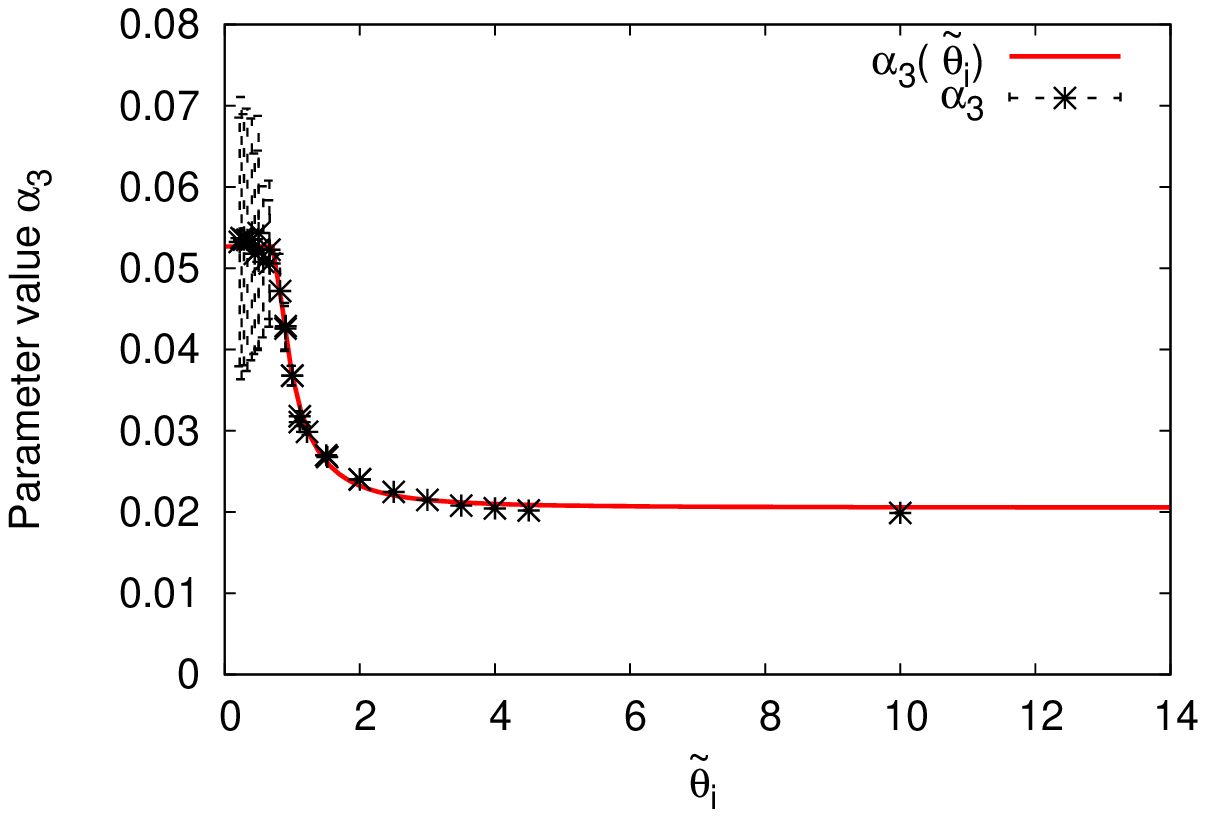}
\end{minipage}
\hfill
 \begin{minipage}{0.49\textwidth}
\includegraphics[width=\textwidth]{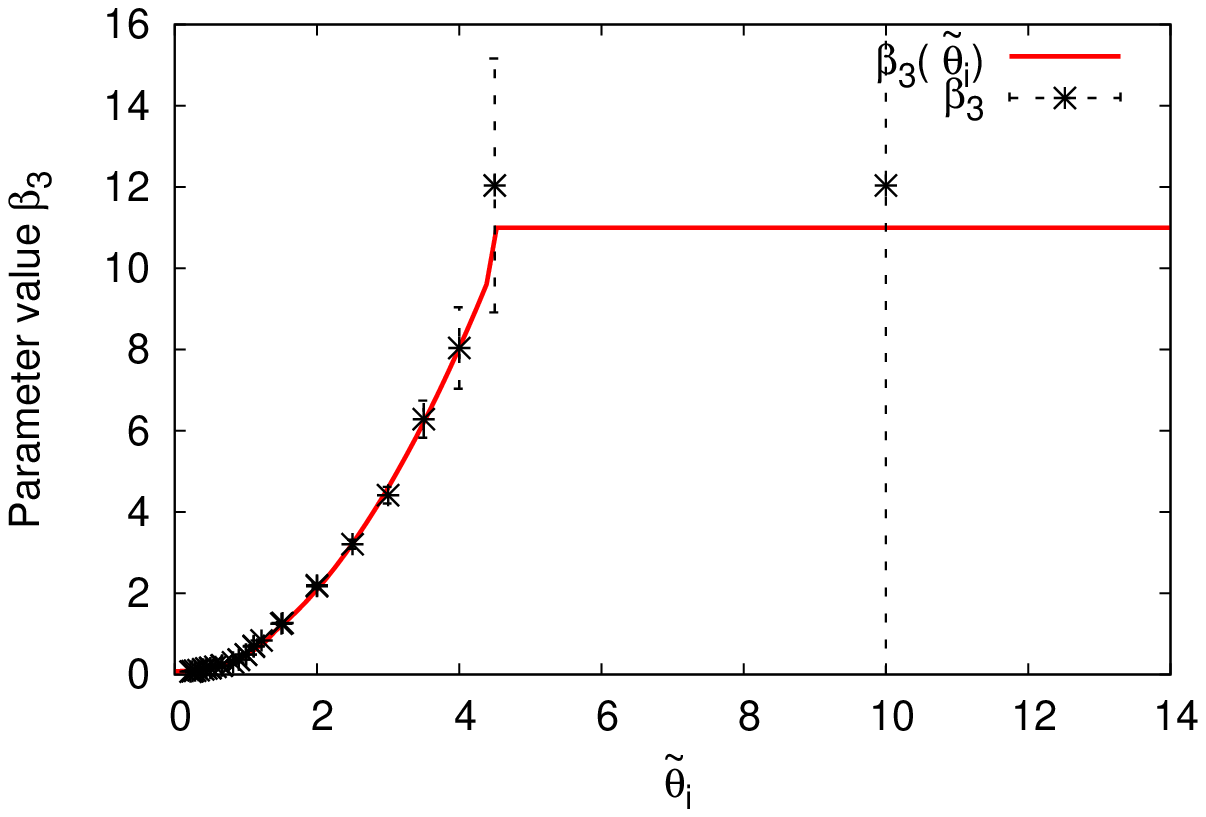}
\end{minipage}
\\
 \begin{minipage}{0.49\textwidth}
\includegraphics[width=\textwidth]{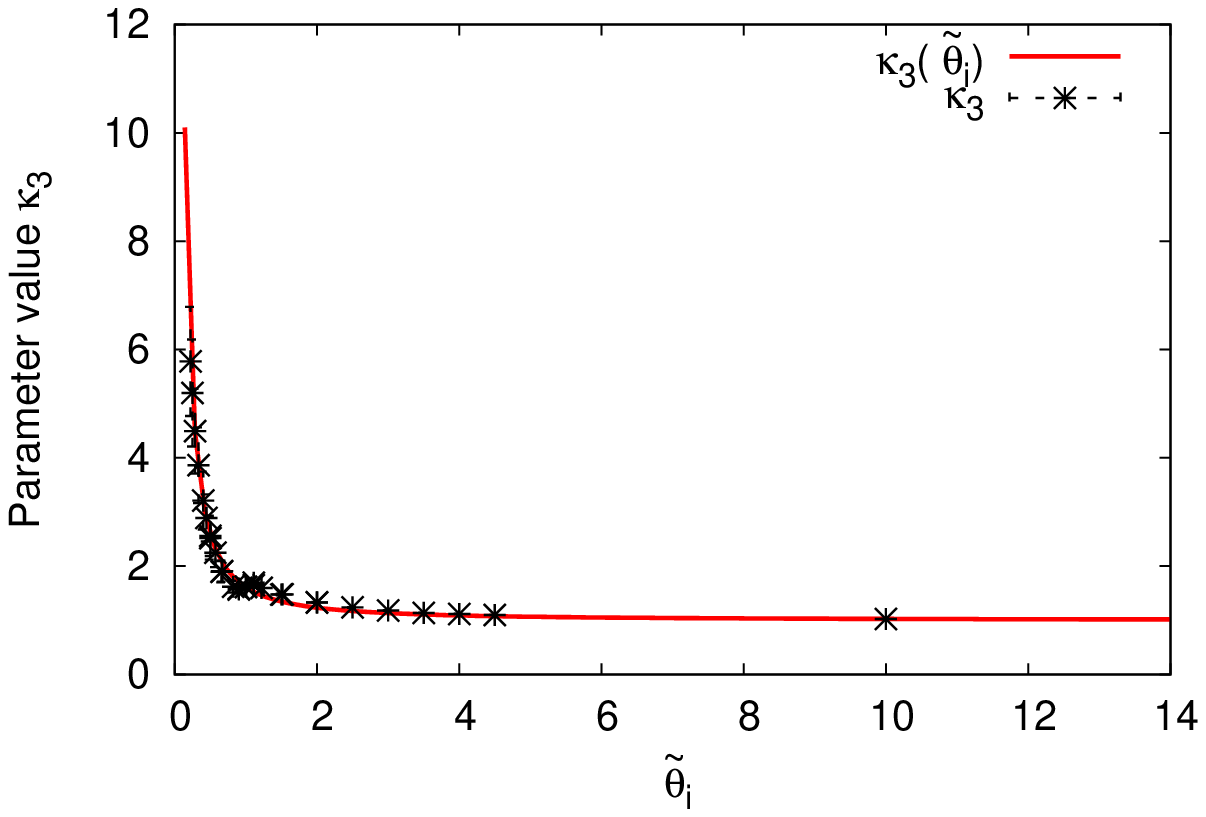}
\end{minipage}
\caption{The parameters $\alpha_{3}$, $\beta_{3}$, and $\kappa_{3}$ of expression $\eqref{eq:f3}$ as a function of $\TTi$ (black $\mathrlap{+}\times$) where the error bars denotes the uncertainty on the parameter in the fits to \eqref{eq:f3} along with the power law fits to \eqref{eq:parameters} (red line).}
\label{fig:para}
\end{figure}

The value of the fitting parameters $\alpha_3$, $\beta_3$ and $\kappa_3$ varies with $\TTi$. In order to make an algorithm that provides $P(\TTf)$ for any given $\TTi$, the $\TTi$ dependence of the parameters $\alpha_3$, $\beta_{3}$ and $\kappa_{3}$ was each fitted to an effective model that could approximate the numerical results. This turned out to be: 
\begin{align} 
&\alpha_3(\TTi)= \begin{cases}
a_{1} & \text{for $\TTi<0.78$}\\
a_{2}\TTi^{- a_{3}}+ a_{4}& \text{otherwise} \nonumber \\
\end{cases}\\
&\beta_3(\TTi) = \begin{cases}
b_{1} \TTi^{b_{2}}\;\;\;+ b_{3} & \text{for $\TTi<1.38$}\\
b_{4} \TTi^{b_{5}}\;\;\;+ b_{6} & \text{for $1.38<\TTi<4.5$} \nonumber \\
b_{7} & \text{otherwise}
\end{cases} \\
&\kappa_{3}(\TTi)=\;\;\;k_{1} \TTi^{k_{2}}\;\;\;+1
\label{eq:parameters}
\end{align}
The results for each expression in \eqref{eq:parameters} can be seen in Fig. \ref{fig:para} and the corresponding parameters in table \ref{tab:param}. 
\begin{table}[ht]
\centering
\caption{The parameters of the new waviness model used in eq. \eqref{eq:f3} and \eqref{eq:parameters}.}
\label{tab:param}
\begin{tabular}{lccccccc}
\hline\hline 
$\alpha_3(\TTi) $& $a_{1}$ & $a_{2} $& $a_{3} $& $a_{4}$ &&& \\ 
		& 0.0527 & 0.0162 & 2.6 & 0.0205 \\
\hline\hline
$\beta_3(\TTi) $&$ b_{1} $&$ b_{2} $&$ b_{3} $&$ b_{4} $&$ b_{5} $&$ b_{6} $&$ b_{7}$\\ 
		&  0.395 & 2.5 & 0.076 & 0.541 & 1.9 & 0.007  & 11 \\
\hline\hline
$\kappa_3(\TTi) $&$k_{1} $&$ k_{2} $ &&&&&\\ 
	& 0.61 &1.39 \\
	\hline\hline
\end{tabular}
\end{table}

\begin{figure}[h!]
\centering
\begin{subfigure}[b]{0.49\textwidth}
\includegraphics[width=\textwidth]{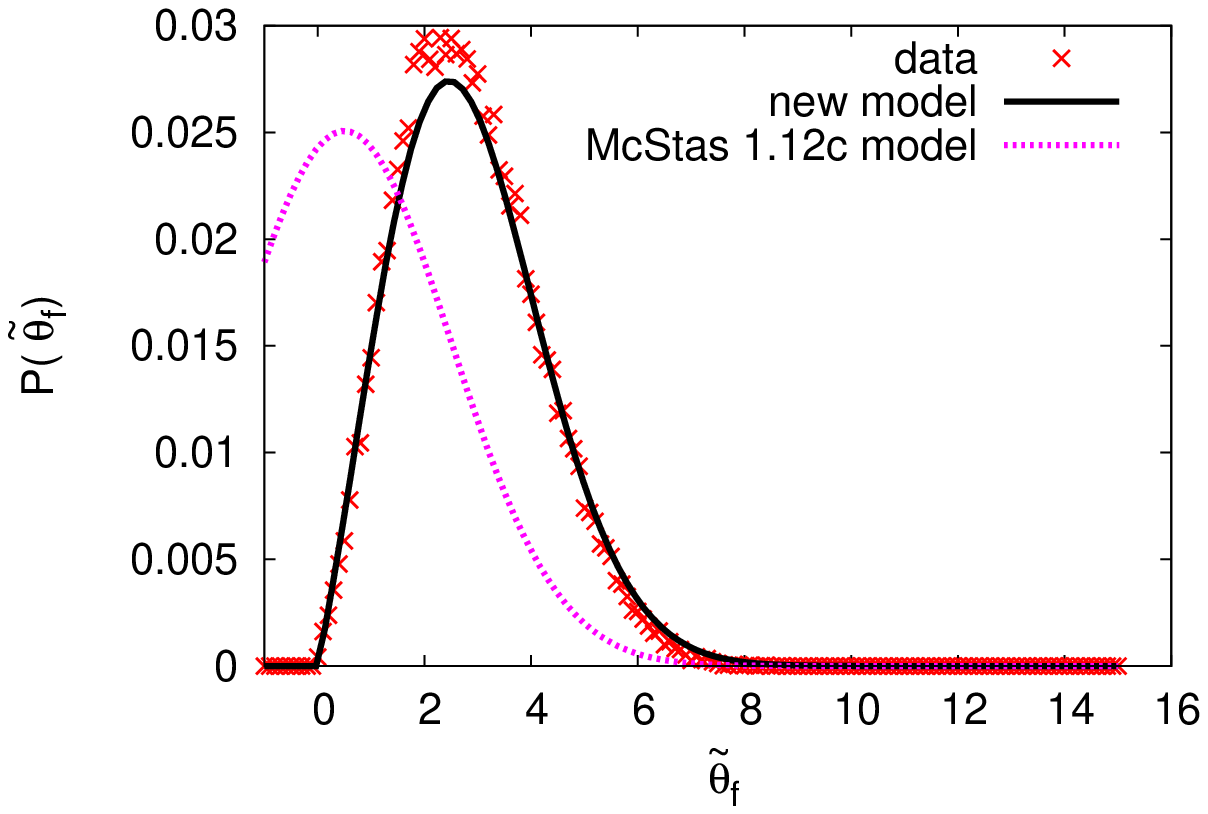}
\end{subfigure}
\hfill
\begin{subfigure}[b]{0.49\textwidth}
\includegraphics[width=\textwidth]{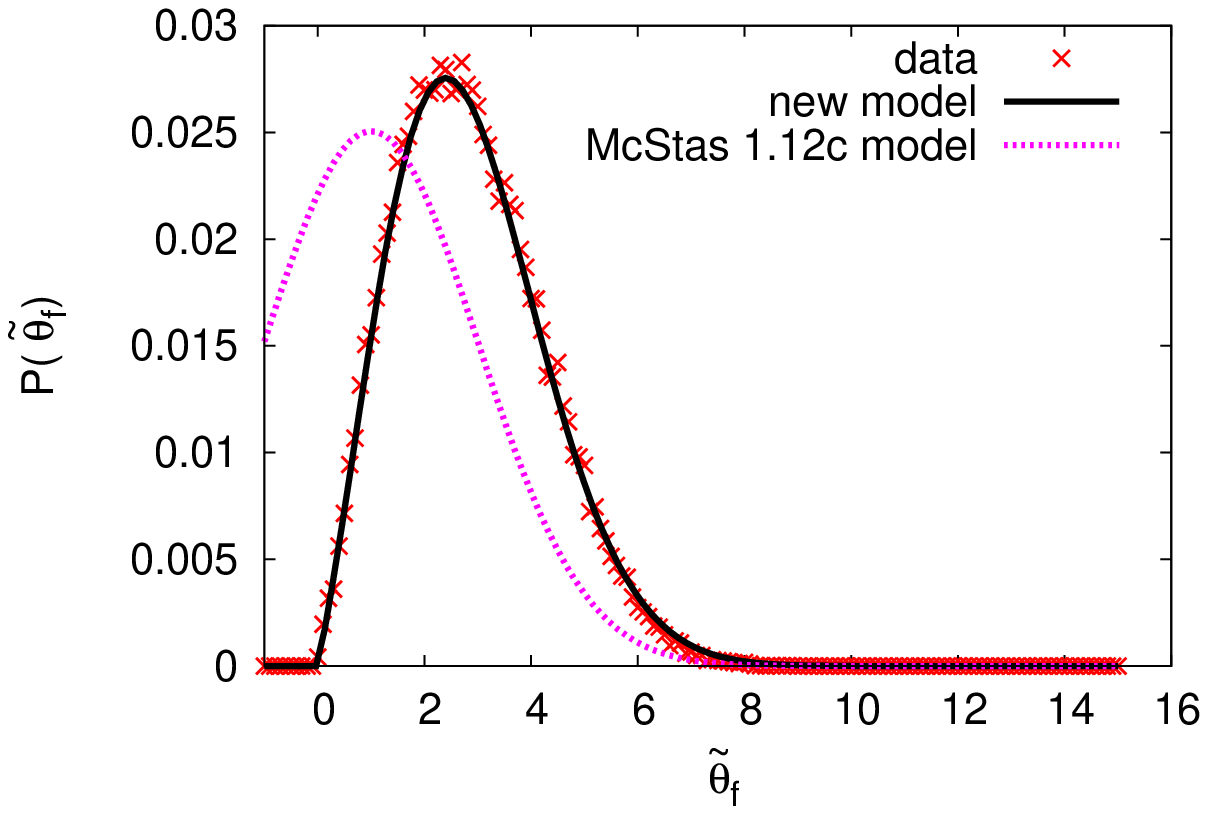}
\end{subfigure}
\\
\begin{subfigure}[b]{0.49\textwidth}
\includegraphics[width=\textwidth]{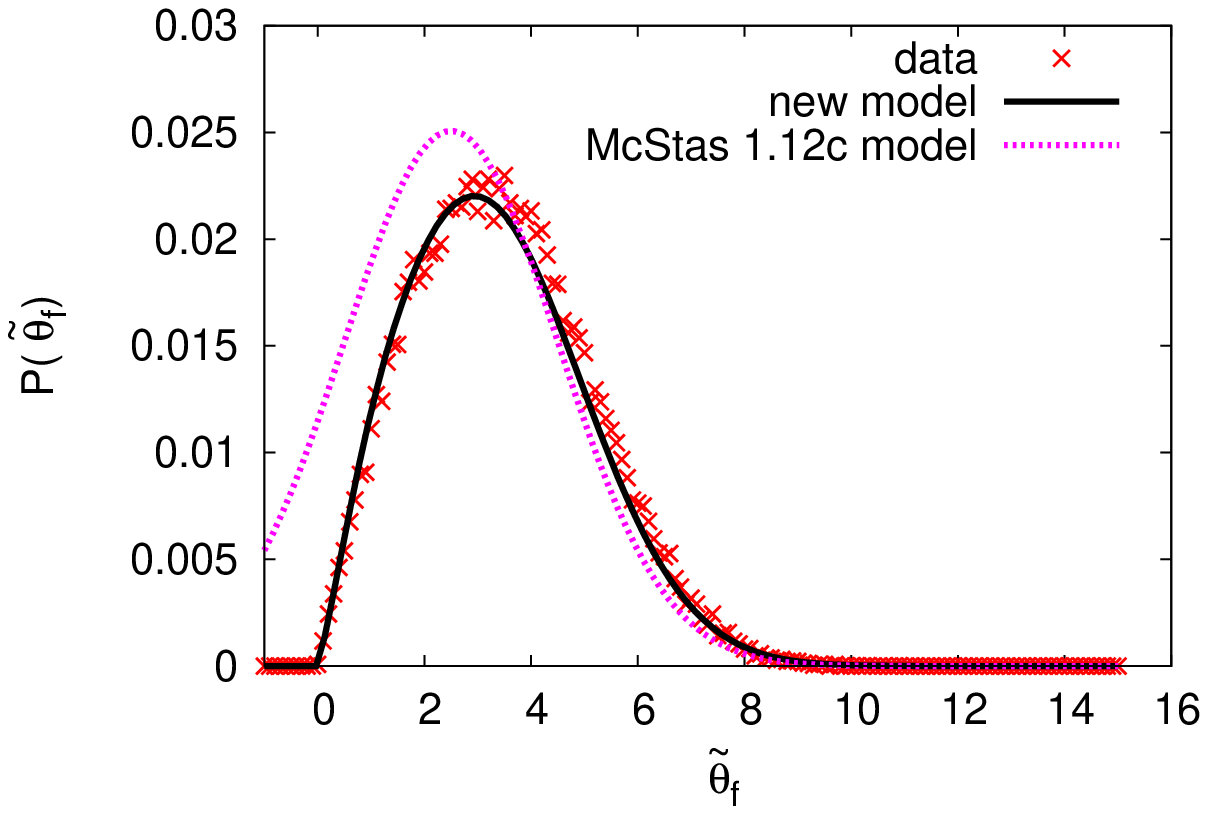}
\end{subfigure}
\hfill
\begin{subfigure}[b]{0.49\textwidth}
\includegraphics[width=\textwidth]{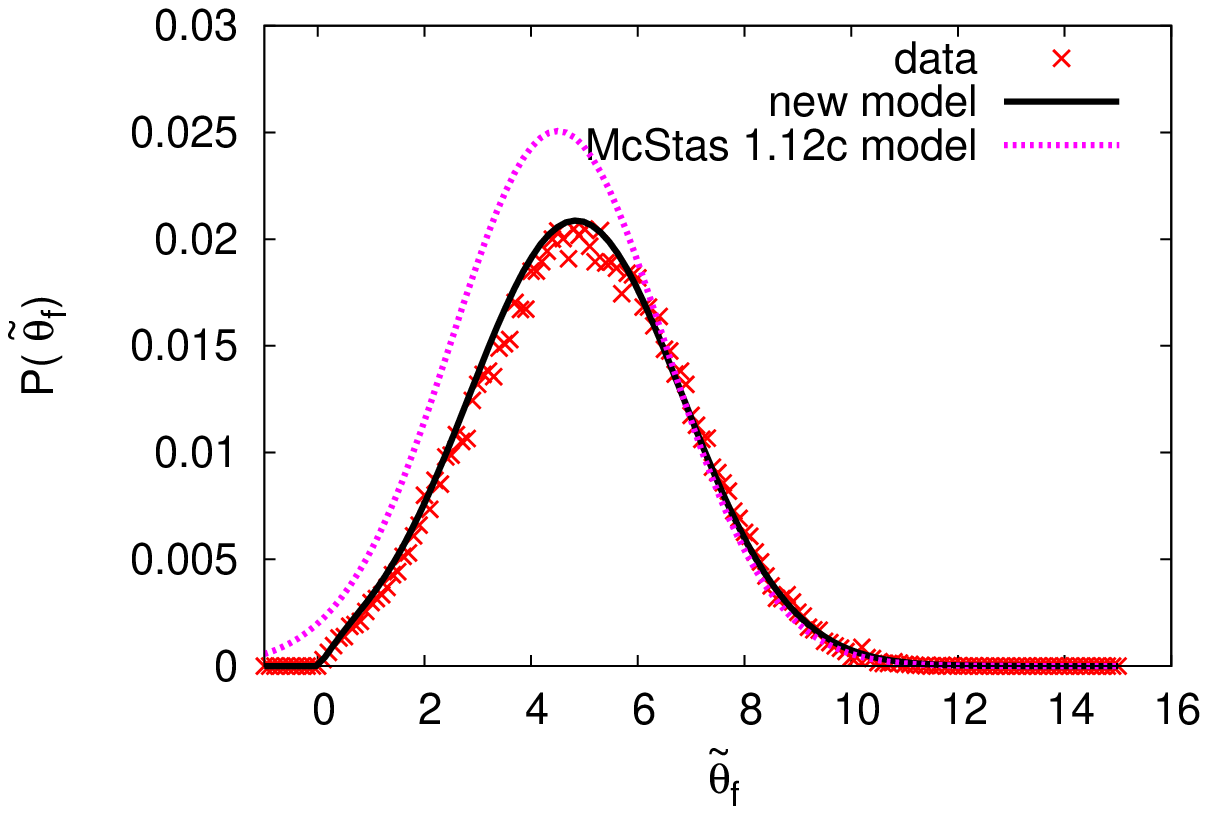}
\end{subfigure}
\caption{Simulated data for $\TTi =0.22$, 1, 2.5 and 4.5 with $N=5\cdot10^4$ rays along with the model of McStas 1.12c (magenta dashed line) and the prediction by the new algorithm (black line).}
\label{fig:modeltest}
\end{figure}

The validity of this model is tested by its strength to predict the simulations made in section \ref{sec:sim}. In Fig. \ref{fig:modeltest} the new model is compared to that of McStas 1.12c for $\TTi =0.22$, 1, 2.5 and 4.5. The results  from the new model over the whole range are sufficiently close to the simple ray tracing simulations that we do not need to refine the model further. 

It should be noted that in the McStas 1.12c model, described in section \ref{sec:oldwavy}, the parameter $w$ was taken to be the width of the distribution of final angles. In the new model $w$ is the average inclination (rms) of the surface, which gives a width of $2w$. We have corrected for this in our comparison in Fig. \ref{fig:modeltest}.
\FloatBarrier
\section{Implementation of waviness in McStas}
\begin{figure}[h!]
\centering
 \begin{minipage}{0.49\textwidth}
\includegraphics[width=\textwidth]{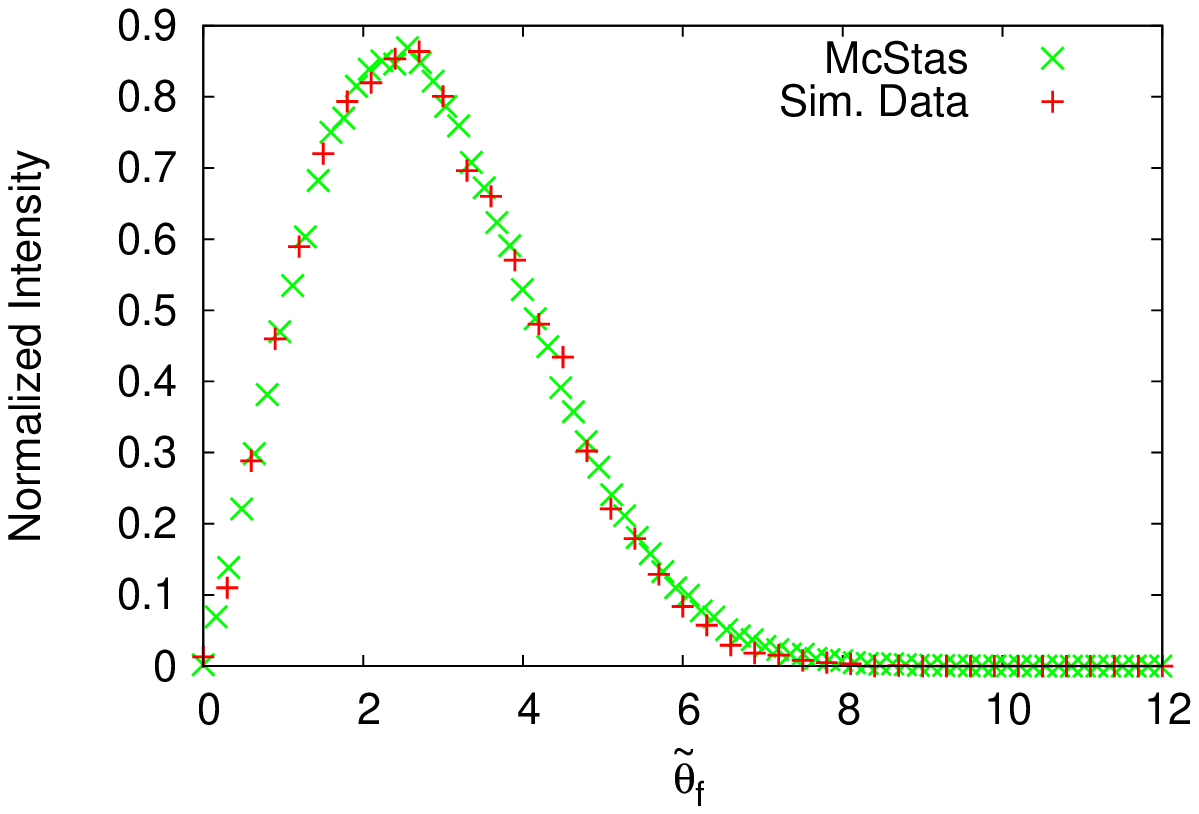}
\end{minipage}
\hfill
 \begin{minipage}{0.49\textwidth}
\includegraphics[width=\textwidth]{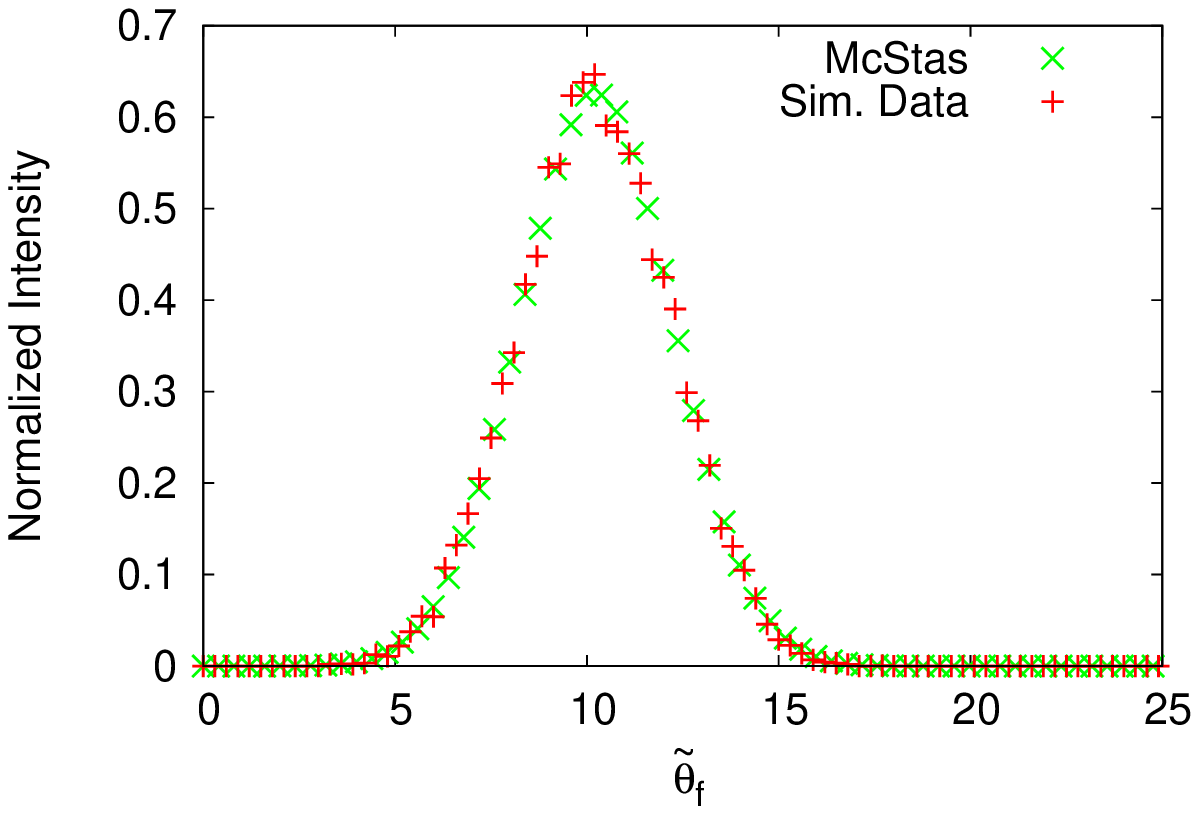}
\end{minipage}
\caption{Simulated data with $N=5\cdot10^4$ rays (red) and normalized neutron counts from McStas (red) made with $N=10^6$ rays for $\TTi =1$ (left) and $\TTi =10$ (right).}
\label{fig:fit_ex}
\end{figure}

The algorithm describing $P(\TTf)$ from eq.~\eqref{eq:f3} along with a {\it hit and miss}\cite{cowan_statistical_1998} sampling routine, were implemented in a straight guide in McStas. First the outcome of a single reflection from a wavy surface for an extremely narrow and well collimated beam was compared to the simulations made in section \ref{sec:sim}. Fig. \ref{fig:fit_ex} shows examples for $\TTi=1$ and $\TTi=10$. The implementation of waviness in McStas clearly reproduces the simulations from section \ref{sec:sim}.
%%%
%%%
\subsection{Example guide simulations}
At last, we show simulations of a 150 m long straight guide with a cross section of $0.05\times0.05\,\mathrm{m^{2}}$ starting 4~m from a $0.1\times0.1\,\mathrm{m^{2}}$ moderator using only 4 \AA\ neutrons. The guide has a coating described by \eqref{eq:ref} with $m=2$, $Q_{\rm c}=0.0217$, $R_{0} =0.99$, $W=0.003$ and $\alpha=2.0$. Simulations were done without waviness, with the McStas 1.12c waviness description (see section \ref{sec:oldwavy}) and with the new description developed in section \ref{sec:newwavy}, respectively. %
\begin{figure}[h!]
\centering
 \begin{minipage}{0.49\textwidth}
\includegraphics[width=\textwidth]{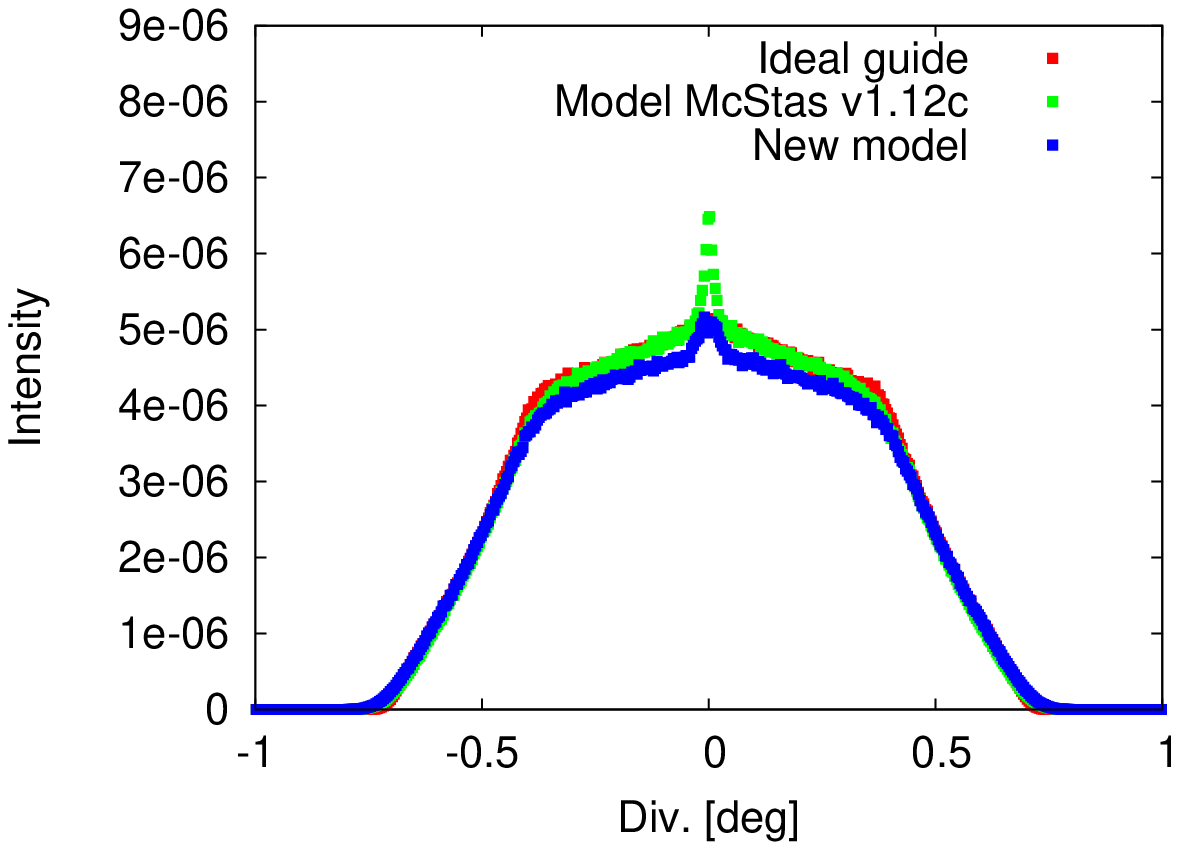}
\end{minipage}
\hfill
 \begin{minipage}{0.49\textwidth}
\includegraphics[width=\textwidth]{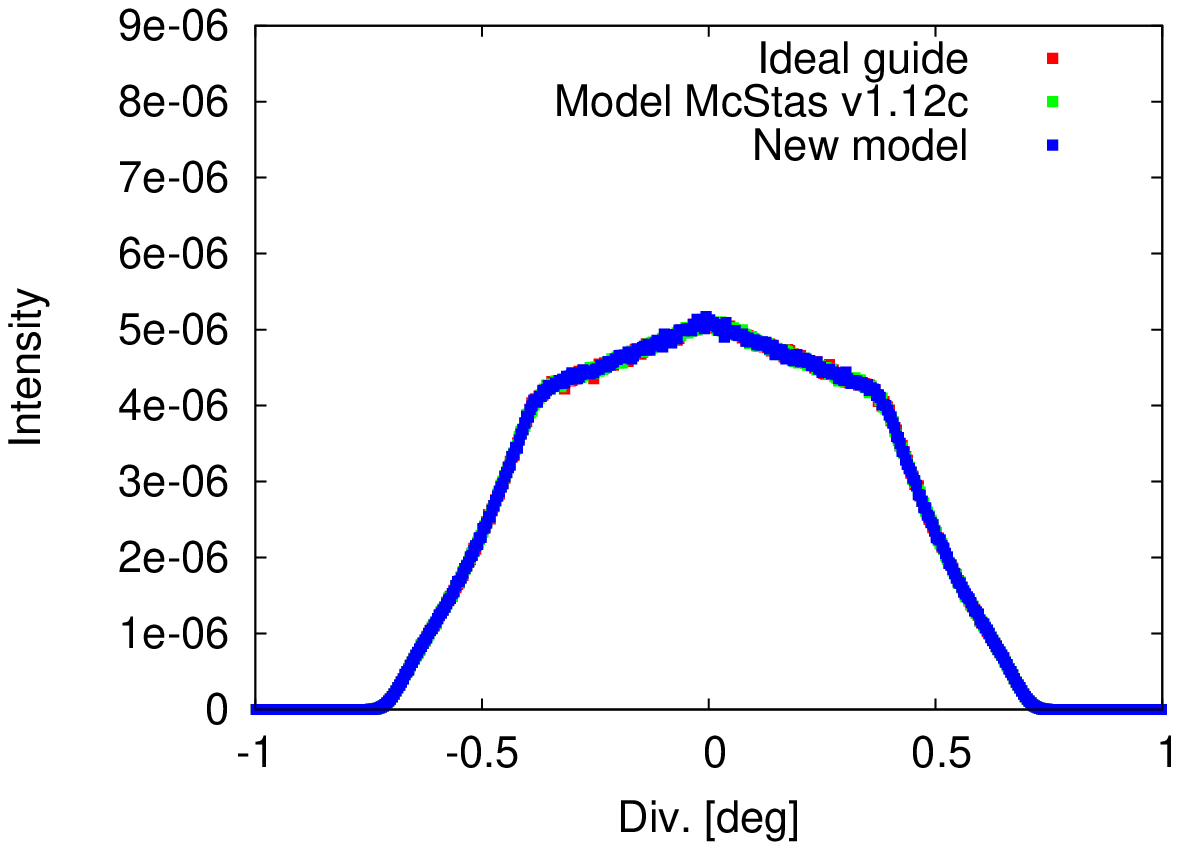}
\end{minipage}
\caption{Intensity as a function of divergence after an ideal guide without waviness (red), with the McStas 1.12c waviness description (green) and with the new waviness description (blue) for $w=10^{-4}$ (left) and $w=10^{-6}$ rad (right). The simulations were made with $10^{7}$ neutron rays each.}
\label{fig:wavyguide}
\end{figure}

When simulating long neutron guides, there is a significant difference between the new and the McStas 1.12c waviness descriptions for large waviness values, $w=10^{-4}$ rad, as shown in Fig. \ref{fig:wavyguide}. In the old version we see a large unphysical peak around $0^{\circ}$ divergence which would lead to a brilliance transfer (defined as neutron flux within a small $2D$ divergence interval and wavelength interval) larger than unity. This, in turn is a violation the Liouville theorem\cite{liouville}. The reason for this is the oversampling of low values of $\theta_f$ caused by a mistake in the algorithm. 

The main consequence of waviness is in the case of the new description reduced intensity as expected, but no violation of the Liouville theorem. The difference between the two descriptions diminishes for smaller values of $w$, an example with a very low waviness $w=10^{-6}$ rad is also shown in Fig. \ref{fig:wavyguide}. 
\section{Summary}
We have performed analytical and simple ray-tracing analysis of the waviness problem relevant for simulations of neutron supermirror reflectivity. The simulations provided a distribution of outgoing angles for a given incoming angle and waviness. It was found that the shape of the distribution evolved as a function of incoming angle. 

For each incident angle the distribution was fitted to a expressions whose parameters was taken to be dependent on the incoming angle. This dependence was fitted as well, which resulted in an effective description of the simulated probability distribution of outgoing angles as a function of the incoming angles and waviness. 

A routine sampling of the outgoing angle for a given incoming angle and waviness was implemented in a straight-guide component in McStas. There is good accordance between the McStas simulations of a single reflection on a wavy surface of a narrow beam and the ray-tracing simulations described in this work.

The unphysical behavior of the McStas 1.12c waviness model for large waviness values is no longer present in the new model.  Instead, the main consequence of waviness is a reduced intensity as expected from simple physical arguments.

\section*{Acknowledgements}
This project has been funded by the University of Copenhagen 2016 program through the project CoNeXT. CoNeXT is a University of Copenhagen interfaculty collaborative project, which is fertilising the ground and harvesting the full potential of the new neutron and X-ray research infrastructures close to Copenhagen University. We also thank the Danish Agency for Research and Innovation for their support through the contribution to the ESS design update phase.

\bibliographystyle{plainnat}
\bibliography{bibtex}

\end{document}